\documentclass[12pt, onecolumn]{IEEEtran}
\renewcommand{\baselinestretch}{1.5}

\usepackage{graphicx}

\newtheorem{theorem}{\bf Theorem}
\newtheorem{corollary}{\bf Corollary}

\newtheorem{lemma}{\bf Lemma}

\newtheorem{definition}{\it Definition}

\begin{document}
%
\title{Error Performance of Channel Coding in Random Access Communication}
%
%
\author{Zheng Wang, \IEEEmembership{Student Member, IEEE}, Jie Luo, \IEEEmembership{Member, IEEE}
\thanks{The authors are with the Electrical and Computer Engineering Department, Colorado State University, Fort Collins, CO 80523. E-mail: \{zhwang, rockey\}@engr.colostate.edu. }
\thanks{This work was supported by the National Science Foundation under Grants CCF-0728826 and CCF-1016985.}
}

\maketitle

\begin{abstract}
A new channel coding approach was proposed in \cite{ref Luo09} for random multiple access communication over the discrete-time memoryless channel. The coding approach allows users to choose their communication rates independently without sharing the rate information among each other or with the receiver. The receiver will either decode the message or report a collision depending on whether reliable message recovery is possible. It was shown that, asymptotically as the codeword length goes to infinity, the set of communication rates supporting reliable message recovery can be characterized by an achievable region which equals Shannon's information rate region possibly without a convex hull operation. In this paper, we derive achievable bounds on error probabilities, including the decoding error probability and the collision miss detection probability, of random multiple access systems with a finite codeword length. Achievable error exponents are obtained by taking the codeword length to infinity.
\end{abstract}


%
\IEEEpeerreviewmaketitle


\section{Introduction}
\label{SectionI}
In multiple access communication, two or more users (transmitters) send messages to a common receiver. The transmitted messages confront distortion both from channel noise and from multi-user interference. Two related communication models, the multi-user information theoretic model and the random access model, have been intensively studied in the literature \cite{ref Gallager85}.

Information theoretic multiple access model, on one hand, assumes each user is backlogged with an infinite reservoir of traffic. Users should first jointly determine their codebooks and information rates, and then send the encoded messages to the receiver continuously over a long communication duration. The only responsibility of the receiver is to decode the messages with its best effort. Under these assumptions, channel capacity and coding theorems are proved by taking the codeword length to infinity \cite{ref Shannon48}\cite{ref Cover05}. Rate and error performance tradeoffs of single user and multiple access systems were analyzed in \cite{ref Gallager65}\cite{ref Gallager85}. Information theoretic model uses symbol-based statistics to characterize the communication channel. Such a physical layer channel model enables rigorous understandings about the impact of channel noise and multi-user interference. However, classical coding results have been derived under the assumption of coordinated communication, in the sense of joint codebook and information rate determination among the multiple users and the receiver. Such an assumption precludes the common scenarios of short messages and bursty traffic arrivals, since in these cases the overhead of full communication coordination is often expensive or infeasible.

Random multiple access model, on the other hand, assumes bursty message arrivals. According to message availability, users independently encode their messages into packets and randomly send these packets to the receiver. It is often assumed that the transmitted packets should be correctly received if the power of the multi-user interference is below a threshold. Otherwise the receiver should report a packet collision and the involved packets are erased \cite{ref Ephremides98}\cite{ref Bertsekas92}. Standard networking regards packet as the basic communication unit, and counts system throughput in packets per time slot as opposed to bits/nats per symbol. Communication channel is characterized using packet-based models, such as the collision channel model \cite{ref Abramson70} and the multipacket-reception channel model \cite{ref Ghez88}\cite{ref Luo06}. Although packet-based models are convenient for upper layer networking \cite{ref Karn90}, their abstract forms essentially prevent an insightful understanding about the impact of physical layer communication to upper layer networking.

In \cite{ref Luo09}, a new channel coding approach was proposed for time-slotted random multiple access communication over a discrete-time memoryless channel using a symbol-based physical layer channel model. Assume in each time slot, each user independently encodes an arbitrary number of data units into a packet and transmits the packet to the receiver. Define the normalized number of data units per symbol as the communication rate of a user in a time slot, which is shared neither among the users nor with the receiver. It was shown in \cite{ref Luo09} that, fundamental performance limitation of the random multiple access system can be characterized using an achievable rate region in the following sense. As the codeword length goes to infinity, if the random communication rate vector of the users happens to be inside the rate region, the receiver can decode all messages with zero asymptotic error probability; if the random communication rate vector happens to be outside the rate region, the receiver can detect a packet collision with an asymptotic probability of one. The achievable rate region was shown to equal to Shannon's information rate region, possibly without a convex hull operation.

In this paper, we derive stronger versions of the coding theorems given in \cite{ref Luo09}  by characterizing the achievable rate and error performance of random multiple access communication over a discrete-time memoryless channel with a {\it finite} codeword length. Our work is motivated by the existing non-asymptotic channel coding results, surveyed in \cite{ref Polyanskiy10}, for classical single-user communication. Following the framework of \cite{ref Luo09}, we assume the random multiple access system predetermines an ``operation region" of the rate vectors in the following sense. For all communication rate vectors within the region, the system {\it intends} to decode the messages; while for all communication rate vectors outside the region, the system {\it intends} to report a packet collision. Given the operation region, there are two types of error events. If the communication rate vector is within the region, the event that the receiver fails to decode the messages correctly is defined as a decoding error event. If the communication rate vector is outside the region, the event that the receiver fails to report a collision is defined as a collision miss detection event. An achievable bound on the system error probability, defined as the maximum of the decoding error probability and the collision miss detection probability, is obtained under the assumption of a finite codeword length. We show that, if the operation region is strictly contained in an achievable rate region, then the system error probability can decrease exponentially in the codeword length. The corresponding exponent is defined as the system error exponent, whose achievable bound is obtained from the error probability bound by taking the codeword length to infinity.

The rest of the paper is organized as follows. With a practical definition of communicate rate, we investigate the error performance of singer-user and multi-user random access systems in Sections \ref{Section SingleUser} and \ref{Section MultipleUser}, respectively. The results are then extended in Section \ref{Section StandardRate} to systems with generalized random coding schemes using the standard communication rate definition, originally introduced in \cite{ref Luo09}. Further discussions and conclusions are provided in Section \ref{Section Conclusion}.

\section{Rate and Error Performance of Single-user Random Access Communication}
\label{Section SingleUser}
For easy understanding, we will first consider single-user random access communication over a discrete-time memoryless channel. The channel is modeled by a conditional distribution function $P_{Y|X}$, where $X\in {\cal X}$, $Y\in {\cal Y}$ are the channel input and output symbols, ${\cal X}$, ${\cal Y}$ are the finite input and output alphabets, respectively. Assume time is partitioned into slots each equaling $N$ symbol durations, which is also the length of a packet. As in \cite{ref Luo09}, we focus on coding within a time slot or a packet.

Suppose the transmitter has no channel information except knowing the channel alphabets\footnote{The significance of this assumption will become clear when we investigate multi-user systems.}. At the beginning of each time slot, according to message availability and the MAC layer protocol, the transmitter chooses a communication rate $r \in \{r_1, \cdots, r_M\}$ without sharing this rate information with the receiver. Here $\{r_1, \cdots, r_M\}$ is a predetermined set of rates, in nats per symbol, with cardinality $M$, known by both the transmitter and the receiver. The transmitter then encodes $\lfloor Nr \rfloor$ data nats, denoted by a message $w$, into a codeword using a ``random coding scheme'' described as follows \cite{ref Luo09}\footnote{Note that the coding scheme is an extended version of the random coding introduced in \cite{ref Shamai07}.}. Let ${\cal L} = \{{\cal C}_{\theta}: \theta \in \Theta \}$ be a library of codebooks indexed by a set $\Theta$. Each codebook contains $M$ classes of codewords. The $i^{th}$ ($i\in \{1, \cdots, M\}$) codeword class contains $\lfloor e^{Nr_i} \rfloor$ codewords, each of $N$ symbol length. Let $\mathcal{C}_{\theta}(w,r)_j$ be the $j^{th}$ codeword symbol of message and communication rate pair $(w,r)$ in codebook $\mathcal{C}_{\theta}$, for $j \in \{1, \cdots, N\}$. The transmitter first randomly generates $\theta$ according to a distribution $\gamma$, such that random variables $X_{(w,r),j}: \theta \rightarrow \mathcal{C}_{\theta}(w,r)_j$ are independently distributed according to an input distribution $P_{X|r}$\footnote{We allow the input distribution to be a function of communication rate. In other words, codewords corresponding to different communication rates may be generated according to different input distributions.}. The random access codebook $\mathcal {C}_{\theta}$ is then used to map the message into a codeword. This is equivalent to mapping a message and rate pair $(w, r)$ into a codeword, denoted by $\mbox{\boldmath$x$}_{(w, r)}$, of $N$ channel input symbols.

We assume the receiver knows the channel $P_{Y|X}$ and the randomly generated codebook $\mathcal{C}_{\theta}$\footnote{This can be realized by sharing the codebook generation algorithm with the receiver.}. Based on this information, the receiver chooses a rate subset $\mathcal{R}\subseteq\{r_1, \cdots, r_M\}$. According to the channel output symbol vector $\mbox{\boldmath $y$}$, the receiver outputs an estimated message and rate pair $(\hat{w}, \hat{r})$ if and only if $\hat{r}\in \mathcal{R}$ and a predetermined decoding error probability requirement is satisfied. Otherwise the receiver outputs a collision. Note that the term ``collision" here is used to maintain consistency with the networking terminology. Throughout the paper, collision means outage, irrespective whether it is caused by multi-user interference or by excessive channel noise.

Since the receiver {\it intends} to decode all messages with $r \in \mathcal{R}$ and to report collision for messages with $r \not\in \mathcal{R}$, we say $\mathcal{R}\subseteq\{r_1, \cdots, r_M\}$ is the ``operation region" of the system. Conditioned on $(w, r)$ is transmitted, for $r\in \mathcal{R}$, we define the decoding error probability as
\begin{equation}\label{DecodeErrorPro}
P_e(w, r)=Pr\{(\hat{w}, \hat{r})\ne (w, r)|(w, r)\}, \qquad \forall (w, r), r \in \mathcal{R}.
\end{equation}
For $r \not\in \mathcal{R}$, we define the collision miss detection probability as
\begin{equation}\label{CollisionPro}
\bar{P}_c(w, r)=1-Pr\{\mbox{``collision"} | (w, r)\}, \qquad \forall (w, r), r \not\in \mathcal{R}.
\end{equation}
Assume $r< I_r(X; Y)$ for all $r \in \mathcal{R}$, where $I_r(X; Y)$ is the mutual information between $X$ and $Y$ computed using input distribution $P_{X|r}$. According to \cite{ref Luo09}, we have the following asymptotic results,
\begin{eqnarray}\label{LimitPr}
&&\lim_{N\to\infty}P_e(w, r)=0, \quad \forall (w, r), r\in \mathcal{R}, \nonumber \\
&&\lim_{N\to\infty}\bar{P}_c(w, r)=0, \quad \forall (w, r), r \not\in \mathcal{R}.
\end{eqnarray}
In other words, asymptotically, the receiver can reliably decode the message if the random communication rate $r$ is inside the operation region; the receiver can reliably report a ``collision'' if $r$ is outside the operation region.

Equation (\ref{LimitPr}) only gives the asymptotic limits on the error probabilities. In the rest of this section, we derive an achievable error probability bound under the assumption of {\it finite} codeword length $N$.

Define the system error probability $P_{es}$ as
\begin{equation}
P_{es}=\max\left\{ \max_{(w, r), r\in \mathcal{R}}P_e(w, r), \max_{(w, r), r\not\in \mathcal{R}} \bar{P}_c(w, r) \right\}.
\end{equation}
The following theorem gives an achievable upper bound on $P_{es}$.
\begin{theorem}{\label{Theorem1}}
Consider single-user random access communication over discrete-time memoryless channel $P_{Y|X}$. Assume random coding with input distributions $P_{X|r}$, defined for all $r\in \{r_1, \cdots, r_M\}$. Let $\mathcal{R}\subseteq \{r_1, \cdots, r_M\}$ be an operation region. Given a codeword length $N$, there exists a decoder whose system error probability $P_{es}$ is upper bounded by
\begin{equation}
P_{es}\le \max\left\{\begin{array}{l} \max_{r\in \mathcal{R}}\left[\sum_{\tilde{r} \in \mathcal{R}} \exp\{-NE_m(\tilde{r}, P_{X|r}, P_{X|\tilde{r}})\}+ \max_{\tilde{r}\not\in \mathcal{R}} \exp\{-NE_i(r, P_{X|r}, P_{X|\tilde{r}})\}\right], \\ \sum_{r\in \mathcal{R}}\max_{\tilde{r}\not\in \mathcal{R}} \exp\{-NE_i(r, P_{X|r}, P_{X|\tilde{r}})\} \end{array} \right\},
\label{SingleUserSystemErrorBound}
\end{equation}
where $E_m(\tilde{r}, P_{X|r}, P_{X|\tilde{r}})$ and $E_i(r, P_{X|r}, P_{X|\tilde{r}})$ are given by
\begin{eqnarray}
&& E_m(\tilde{r}, P_{X|r}, P_{X|\tilde{r}})=\max_{0<\rho\le 1}-\rho \tilde{r} + \max_{0<s\le 1}-\log \sum_Y\left[\sum_X P_{X|r}(X)P(Y|X)^{1-s}\right]\left[\sum_X P_{X|\tilde{r}}(X)P(Y|X)^{\frac{s}{\rho}}\right]^{\rho}, \nonumber \\
&& E_i(r, P_{X|r}, P_{X|\tilde{r}})=\max_{0<\rho\le 1} -\rho r +\max_{0<s\le 1-\rho} -\log \sum_Y\left[ \sum_X P_{X|r}(X)P(Y|X)^{\frac{s}{s+\rho}}\right]^{s+\rho} \nonumber \\
&& \qquad \qquad \times \left[\sum_X P_{X|\tilde{r}}(X)P(Y|X)\right]^{1-s}.
\label{EmEiSingle}
\end{eqnarray}
$\QED$
\end{theorem}

The proof of Theorem \ref{Theorem1} is given in Appendix \ref{ProofTheorem1}\footnote{Even though Theorem \ref{Theorem1} is implied by Theorem \ref{Theorem2} given in Section \ref{Section MultipleUser}, we still provide its full proof because it is much easier to follow than the proof of Theorem \ref{Theorem2}. Indeed, we suggest readers should understand the basic ideas in the proof of Theorem \ref{Theorem1} before reading the more sophisticated proof of Theorem \ref{Theorem2}.}. In the proof, we assumed the following decoding algorithm at the receiver to achieve the error probability bound given in (\ref{SingleUserSystemErrorBound}). Upon receiving the channel output symbols $\mbox{\boldmath $y$}$, the receiver outputs an estimated message and rate pair $(w, r)$ with $r\in \mathcal{R}$ if both the following two conditions are satisfied,
\begin{eqnarray}\label{DecodeConditions}
&&\mbox{C1: }-\frac{1}{N}\log Pr\{\mbox{\boldmath $y$}|\mbox{\boldmath $x$}_{(w, r)}\} < -\frac{1}{N}\log Pr\{\mbox{\boldmath $y$}|\mbox{\boldmath $x$}_{(\tilde{w}, \tilde{r})}\},\mbox{ for all } (\tilde{w},\tilde{r})\neq (w, r) , r,\tilde{r} \in \mathcal{R},\nonumber\\
&&\mbox{C2: }-\frac{1}{N}\log Pr\{\mbox{\boldmath $y$}|\mbox{\boldmath $x$}_{(w, r)}\} < \tau_{r}(\mbox{\boldmath $y$}),
\label{CriteriaSingle}
\end{eqnarray}
where $\tau_{r}(\cdot)$ is a pre-determined function of the channel output $\mbox{\boldmath $y$}$, associated with codewords of rate $r$. We term $\tau_{r}(\cdot)$ a typicality threshold function. If there is no codeword satisfying (\ref{DecodeConditions}), the receiver reports a collision. In other words, the receiver decodes only if the log-likelihood of the maximum likelihood estimation exceeds certain threshold. Note that the random access codebook used to encode the message contains a large number of codewords, but the receiver only searches codewords corresponding to rates inside the operation region.

Define the corresponding exponent as the system error exponent $E_s=\lim_{N\to \infty} -\frac{1}{N}\log P_{es}$. Theorem \ref{Theorem1} implies the following achievable bound on $E_s$.

\begin{corollary}{\label{Corollary1}}
The system error exponent of single-user random access communication given in Theorem \ref{Theorem1} is lower-bounded by
\begin{equation}
E_s =\lim_{N\to \infty} -\frac{1}{N}\log P_{es}\ge \min\left\{\min_{r, \tilde{r} \in \mathcal{R}} E_m(\tilde{r}, P_{X|r}, P_{X|\tilde{r}}), \min_{r\in \mathcal{R}, \tilde{r}\not\in \mathcal{R}} E_i(r, P_{X|r}, P_{X|\tilde{r}})   \right\},
\label{E_s_single}
\end{equation}
where $E_m(\tilde{r}, P_{X|r}, P_{X|\tilde{r}})$ and $E_i(r, P_{X|r}, P_{X|\tilde{r}})$ are defined in (\ref{EmEiSingle}). $\QED$
\end{corollary}

Corollary \ref{Corollary1} is implied by Theorem \ref{Theorem1}. An alternative proof can also be found in \cite{ref Wang10}.

Note that if we define the decoding error exponent $E_d$ and the collision miss detection exponent $E_c$ as
\begin{eqnarray}
&& E_d=\min_{(w, r), r \in \mathcal{R}} \lim_{N\to \infty} -\frac{1}{N}\log P_e(w, r), \nonumber \\
&& E_c=\min_{(w, r), r \not\in \mathcal{R}} \lim_{N\to \infty} -\frac{1}{N}\log \bar{P}_c(w, r) ,
\end{eqnarray}
then the system error exponent equals the minimum of the two exponents, i.e., $E_s=\min\{E_d, E_c\}$. The lower bound of $E_s$ given in (\ref{E_s_single}) is obtained by optimizing the typicality threshold function $\tau_{r}(\cdot)$ as done in the proof of Theorem \ref{Theorem1}. It is easy to see that, for each $\mbox{\boldmath $y$}$, the decoding error exponent $E_d$ increases in $\tau_r(\mbox{\boldmath$y$})$, while the collision miss detection exponent $E_c$ decreases in $\tau_r(\mbox{\boldmath$y$})$. Therefore, $\tau_{r}(\cdot)$ can be used to adjust the tradeoff between $E_d$ and $E_c$.

Also note that the first term on the right hand side of (\ref{E_s_single}) corresponds to the maximum likelihood decoding criterion C1 in (\ref{DecodeConditions}). This term becomes Gallager's random-coding exponent \cite{ref Gallager65} if the input distributions associated to all rates are identical. The second term is due to the typical sequence decoding criterion C2 in (\ref{DecodeConditions}). The two criteria, in conjunction, enabled collision detection at the receiver with a good decoding error performance.

We end this section by pointing out that the probability bound given in (\ref{SingleUserSystemErrorBound}) can be further tightened, especially when the input distributions corresponding to $r\in \mathcal{R}$ are similar to each other. In the special case if the input distributions are identical for all rates, then the term $\sum_{\tilde{r} \in \mathcal{R}} \exp\{-NE_m(\tilde{r}, P_{X|r}, P_{X|\tilde{r}})\}$ in (\ref{SingleUserSystemErrorBound}), which corresponds to the maximum likelihood decoding criterion C1 in (\ref{CriteriaSingle}), can be further improved to Gallager's bound given in \cite{ref Gallager65}\footnote{Specifically, we mean the bound given by (18) in \cite{ref Gallager65} with $R=\frac{1}{N}\log\sum_{\tilde{r}\in \mathcal{R}}e^{N\tilde{r}}$.}. However, in a general case, such improvement makes the error bound less structured comparing to (\ref{SingleUserSystemErrorBound}), and it gives the same error exponent results. Therefore, we choose to skip the detailed discussion in the paper.

\section{Rate and Error Performance of Random Multiple Access Communication}
\label{Section MultipleUser}
In this section, we consider $K$-user time-slotted random multiple access communication over a discrete-time memoryless channel. The channel is modeled by a conditional distribution $P_{Y|X_1, \cdots, X_K}$, where $X_k \in {\cal X}_k$, $k \in \{1, \cdots, K\}$, is the channel input symbol of user $k$ with ${\cal X}_k$ being the the finite input alphabet, and $Y \in {\cal Y}$ is the channel output symbol with ${\cal Y}$ being the finite output alphabet. Assume the slot length equals $N$ symbol durations, which is also the length of a packet. We again focus on coding within one time slot.

Suppose at the beginning of a time slot, each user, say user $k$, chooses an arbitrary communication rate $r_k$, in nats per symbol, and encodes $\lfloor Nr_k \rfloor$ data nats, denoted by a message $w_k$, into a packet of $N$ symbols. Assume $r_k \in \{r_{k_1}, \cdots, r_{k_M}\}$, where $ \{r_{k_1}, \cdots, r_{k_M}\}$ is a predetermined set of rates, with cardinality $M$, known at the receiver. We assume the actual communication rates of the users are shared neither among each other, nor with the receiver. Whether the channel is known at the users (transmitters) is not important at this point. Because the global rate information is not available, an individual user cannot know a priori whether or not its rate is supported by the channel in terms of reliable message recovery. Encoding is done using a random coding scheme described as follows. Let $\mathcal{L}_k = \{\mathcal{C}_{k\theta_k}:\theta_k \in \Theta_k\}$ be a codebook library of user $k$, the codebooks of which are indexed by set $\Theta_k$. Each codebook contains $M$ classes of codewords. The $i^{th}$ codeword class contains $\lfloor e^{Nr_{k_i}}\rfloor$ codewords, each with $N$ symbols. Denote $\mathcal{C}_{k\theta_k}(w_k,r_k)_j$ as the $j^{th}$ symbol of the codeword corresponding to message $w_k$ and communication rate $r_k$ in codebook $\mathcal{C}_{k\theta_k}$. User $k$ first generates $\theta_k$ according to a distribution $\gamma_k$, such that random variables $X_{(w_k,r_k),j}: \theta_k \rightarrow \mathcal{C}_{k\theta_k}(w_k, r_k)_j$ are independently distributed according to an input distribution $P_{X|r_k}$. User $k$ then uses codebook $\mathcal{C}_{k\theta_k}$ to map $(w_k, r_k)$ into a codeword, denoted by $\mbox{\boldmath $x$}_{(w_k, r_k)}$, and sends it to the receiver.

Assume the receiver knows the channel $P_{Y|X_1,\cdots, X_K}$ and the randomly generated codebooks of all users. Based on the channel and the codebook information, the receiver predetermines an ``operation region" $\cal R$, which is a set of communication rate vectors under which the receiver {\it intends} to decode the messages. In each time slot, upon receiving the channel output symbol vector $\mbox{\boldmath $y$}$, the receiver outputs the estimated message and rate vector pair $(\hat{\mbox{\boldmath $w$}}, \hat{\mbox{\boldmath $r$}})$ (that contains the estimates for all users) only if $\hat{\mbox{\boldmath $r$}}\in \cal R$ and a predetermined decoding error probability requirement is satisfied. Otherwise the receiver outputs a collision.

To simplify the notations, we will use bold font vector variables to denote the corresponding variables of multiple users. For example, $\hat{\mbox{\boldmath $w$}}$ denotes the message estimates of all users, $\mbox{\boldmath $r$}$ denotes the communication rates of all users, $\mbox{\boldmath $P$}_{\mbox{\scriptsize \boldmath $X$}|\mbox{\scriptsize \boldmath $r$}}$ denotes the input distributions conditioned on communication rates $\mbox{\boldmath $r$}$, etc. For a vector variable $\mbox{\boldmath $r$}$, we will use $r_k$ to denote the element corresponding to user $k$. Let ${\cal S} \subset \{1, \cdots, K\}$ be an arbitrary subset of user indices. We will use $\mbox{\boldmath $r$}_{\cal S}$ to denote the communication rates of users in ${\cal S}$, and will use $\mbox{\boldmath $w$}_{\bar{\cal S}}$ to denote the messages of users not in ${\cal S}$, etc.

Similar to the single-user system, conditioned on $(\mbox{\boldmath $w$}, \mbox{\boldmath $r$})$ is transmitted, we define the decoding error probability for $(\mbox{\boldmath $w$}, \mbox{\boldmath $r$})$ with $\mbox{\boldmath $r$}\in \cal R$ as
\begin{equation}\label{M-DecodeErrorDef}
P_e(\mbox{\boldmath $w$}, \mbox{\boldmath $r$})=Pr\{(\hat{\mbox{\boldmath $w$}}, \hat{\mbox{\boldmath $r$}}) \ne (\mbox{\boldmath $w$}, \mbox{\boldmath $r$})|(\mbox{\boldmath $w$}, \mbox{\boldmath $r$})\}, \forall (\mbox{\boldmath $w$}, \mbox{\boldmath $r$}), \mbox{\boldmath $r$}\in \cal R.
\end{equation}
We define the collision miss detection probabilities for $(\mbox{\boldmath $w$}, \mbox{\boldmath $r$})$ with $\mbox{\boldmath $r$}\not\in \cal R$ as
\begin{equation}\label{M-CollisionErrorDef}
\bar{P}_c(\mbox{\boldmath $w$}, \mbox{\boldmath $r$}) = 1 - Pr\{\mbox{``collision''} | (\mbox{\boldmath $w$}, \mbox{\boldmath $r$}) \}, \forall (\mbox{\boldmath $w$}, \mbox{\boldmath $r$}), \mbox{\boldmath $r$}\not\in \cal R.
\end{equation}

Assume for all $\mbox{\boldmath $r$}\in \mathcal {R} $ and for all user subset ${\cal S} \subset \{1, \cdots, K\}$, we have $\sum_{k\not\in S} r_k < I_{\mbox{\scriptsize \boldmath $r$}}(\mbox{\boldmath $X$}_{\bar{\cal S}};Y|\mbox{\boldmath $X$}_{\cal S})$, where $I_{\mbox{\scriptsize \boldmath $r$}}(\mbox{\boldmath $X$}_{\bar{\cal S}};Y|\mbox{\boldmath $X$}_{\cal S})$ is the conditional mutual information computed using input distribution $\mbox{\boldmath $P$}_{\mbox{\scriptsize \boldmath $X$}|\mbox{\scriptsize \boldmath $r$}}$. According to the achievable region result given in \cite{ref Luo09}, asymptotically, the receiver can reliably decode the messages for all rate vectors inside $\mathcal{R}$ and can reliably report a collision for all rate vectors outside $\mathcal{R}$. In other words,
\begin{eqnarray}
&& \lim_{N\to \infty} P_e(\mbox{\boldmath $w$}, \mbox{\boldmath $r$})=0, \qquad \forall (\mbox{\boldmath $w$}, \mbox{\boldmath $r$}), \mbox{\boldmath $r$}\in \mathcal{R}, \nonumber \\
&&\lim_{N\to \infty} \bar{P}_c(\mbox{\boldmath $w$}, \mbox{\boldmath $r$})=0, \qquad \forall (\mbox{\boldmath $w$}, \mbox{\boldmath $r$}), \mbox{\boldmath $r$}\not\in \mathcal{R}.
\end{eqnarray}

Define the system error probability $P_{es}$ as
\begin{equation}
P_{es}=\max\left\{\max_{(\mbox{\scriptsize\boldmath $w$}, \mbox{\scriptsize\boldmath $r$}), \mbox{\scriptsize\boldmath $r$} \in \mathcal{R}}P_e( \mbox{\boldmath $w$}, \mbox{\boldmath $r$}), \max_{(\mbox{\scriptsize\boldmath $w$}, \mbox{\scriptsize\boldmath $r$}), \mbox{\scriptsize\boldmath $r$} \not\in \mathcal{R}}\bar{P}_c(\mbox{\boldmath $w$}, \mbox{\boldmath $r$})\right\}.
\end{equation}
The following theorem gives an upper bound on $P_{es}$.

\begin{theorem}{\label{Theorem2}}
For $K$-user random multiple access communication over a discrete time memoryless channel $P_{Y|\mbox{\scriptsize \boldmath $X$}}$. Assume finite codeword length $N$, and random coding with input distribution $\mbox{\boldmath $P$}_{\mbox{\scriptsize \boldmath $X$}|\mbox{\scriptsize \boldmath $r$}}$ for all $\mbox{\boldmath $r$}$ with $r_k\in \{r_{k_1}, \cdots, r_{k_M}\}$, $1\le k\le K$. Let $\mathcal{R}$ be the operation region. There exists a decoding algorithm, whose system error probability $P_{es}$ is upper bounded by
\begin{equation}
P_{es}\le \max\left\{ \begin{array}{l} \max_{\mbox{\scriptsize \boldmath $r$}\in \mathcal{R} }\sum_{\mathcal{S}\subset\{1, \cdots, K\}} \left[ \begin{array}{l} \sum_{\tilde{\mbox{\scriptsize \boldmath $r$}}\in \mathcal{R}, \tilde{\mbox{\scriptsize \boldmath $r$}}_{\mathcal{S}}= \mbox{\scriptsize \boldmath $r$}_{\mathcal{S}} } \exp\{-N E_m(\mathcal{S}, \tilde{\mbox{\boldmath $r$}}, \mbox{\boldmath $P$}_{\mbox{\scriptsize \boldmath $X$}|\mbox{\scriptsize \boldmath $r$}}, \mbox{\boldmath $P$}_{\mbox{\scriptsize \boldmath $X$}|\tilde{\mbox{\scriptsize \boldmath $r$}}})\}  \\   +  \max_{\mbox{\scriptsize \boldmath $r$}'\not\in \mathcal{R}, \mbox{\scriptsize \boldmath $r$}'_{\mathcal{S}}= \mbox{\scriptsize \boldmath $r$}_{\mathcal{S}} } \exp\{-NE_i(\mathcal{S}, \mbox{\boldmath $r$}, \mbox{\boldmath $P$}_{\mbox{\scriptsize \boldmath $X$}|\mbox{\scriptsize \boldmath $r$}}, \mbox{\boldmath $P$}_{\mbox{\scriptsize \boldmath $X$}|\mbox{\scriptsize \boldmath $r$}'}) \} \end{array}  \right],  \\  \max_{\tilde{\mbox{\scriptsize \boldmath $r$}}\not\in \mathcal{R}}\sum_{\mathcal{S}\subset\{1, \cdots, K\}} \sum_{\mbox{\scriptsize \boldmath $r$}\in \mathcal{R}, \mbox{\scriptsize \boldmath $r$}_{\mathcal{S}}=\tilde{\mbox{\scriptsize \boldmath $r$}}_{\mathcal{S}}} \max_{\mbox{\scriptsize \boldmath $r$}'\not\in \mathcal{R}, \mbox{\scriptsize \boldmath $r$}'_{\mathcal{S}}= \tilde{\mbox{\scriptsize \boldmath $r$}}_{\mathcal{S}} } \exp\{ -N E_i(\mathcal{S}, \mbox{\boldmath $r$}, \mbox{\boldmath $P$}_{\mbox{\scriptsize \boldmath $X$}|\mbox{\scriptsize \boldmath $r$}}, \mbox{\boldmath $P$}_{\mbox{\scriptsize \boldmath $X$}|\mbox{\scriptsize \boldmath $r$}'}) \}  \end{array}     \right\},
\label{MultiUserSystemErrorBound}
\end{equation}
where $E_m(\mathcal{S}, \tilde{\mbox{\boldmath $r$}}, \mbox{\boldmath $P$}_{\mbox{\scriptsize \boldmath $X$}|\mbox{\scriptsize \boldmath $r$}}, \mbox{\boldmath $P$}_{\mbox{\scriptsize \boldmath $X$}|\tilde{\mbox{\scriptsize \boldmath $r$}}})$ and $E_i(\mathcal{S}, \mbox{\boldmath $r$}, \mbox{\boldmath $P$}_{\mbox{\scriptsize \boldmath $X$}|\mbox{\scriptsize \boldmath $r$}}, \mbox{\boldmath $P$}_{\mbox{\scriptsize \boldmath $X$}|\mbox{\scriptsize \boldmath $r$}'})$ are given by
\begin{eqnarray}
&& E_m(\mathcal{S}, \tilde{\mbox{\boldmath $r$}}, \mbox{\boldmath $P$}_{\mbox{\scriptsize \boldmath $X$}|\mbox{\scriptsize \boldmath $r$}}, \mbox{\boldmath $P$}_{\mbox{\scriptsize \boldmath $X$}|\tilde{\mbox{\scriptsize \boldmath $r$}}}) = \max_{0<\rho \le 1} -\rho \sum_{k\not\in \mathcal{S}}\tilde{r}_k + \max_{0<s\le 1} -\log \sum_Y \sum_{\mbox{\scriptsize \boldmath $X$}_{\mathcal{S}}} \prod_{k\in \mathcal{S}} P_{X|r_k}(X_k)                    \nonumber \\
&& \quad \times \left(\sum_{\mbox{\scriptsize \boldmath $X$}_{\bar{\mathcal{S}}}}\prod_{k \not\in \mathcal{S}}P_{X|r_k}(X_k)P(Y|\mbox{\boldmath $X$})^{1-s}\right)  \left(\sum_{\mbox{\scriptsize \boldmath $X$}_{\bar{\mathcal{S}}}}\prod_{k \not\in \mathcal{S}}P_{X|\tilde{r}_k}(X_k)P(Y|\mbox{\boldmath $X$})^{\frac{s}{\rho}} \right)^{\rho},                       \nonumber \\
&& E_i(\mathcal{S}, \mbox{\boldmath $r$}, \mbox{\boldmath $P$}_{\mbox{\scriptsize \boldmath $X$}|\mbox{\scriptsize \boldmath $r$}}, \mbox{\boldmath $P$}_{\mbox{\scriptsize \boldmath $X$}|\mbox{\scriptsize \boldmath $r$}'}) = \max_{0<\rho \le 1} -\rho \sum_{k\not\in \mathcal{S}}r_k + \max_{0<s \le 1-\rho} - \log \sum_Y \sum_{\mbox{\scriptsize \boldmath $X$}_{\mathcal{S}}} \prod_{k\in \mathcal{S}} P_{X|r_k}(X_k)               \nonumber \\
&& \quad \times \left(\sum_{\mbox{\scriptsize \boldmath $X$}_{\bar{\mathcal{S}}}}\prod_{k \not\in \mathcal{S}}P_{X|r_k}(X_k)P(Y|\mbox{\boldmath $X$})^{\frac{s}{s+\rho}} \right)^{s+\rho}\left(\sum_{\mbox{\scriptsize \boldmath $X$}_{\bar{\mathcal{S}}}}\prod_{k \not\in \mathcal{S}}P_{X|r'_k}(X_k)P(Y|\mbox{\boldmath $X$})\right)^{1-s}.
\label{EmEiMulti}
\end{eqnarray}
$\QED$
\end{theorem}

The proof of Theorem \ref{Theorem2} is given in Appendix \ref{ProofTheorem2}. In the proof, we assumed the following decoding algorithm at the receiver to achieve the error probability bound given in (\ref{MultiUserSystemErrorBound}). Upon receiving the channel output symbols $\mbox{\boldmath $y$}$, the receiver outputs an estimated message vector and rate vector pair $(\mbox{\boldmath $w$}, \mbox{\boldmath $r$})$ with $\mbox{\boldmath $r$}\in \mathcal{R}$ if both the following two conditions are satisfied.
\begin{eqnarray}
&&\mbox{C1: }-\frac{1}{N}\log Pr\{\mbox{\boldmath $y$}|\mbox{\boldmath $x$}_{(\mbox{\scriptsize \boldmath $w$}, \mbox{\scriptsize \boldmath $r$})}\} < -\frac{1}{N}\log Pr\{\mbox{\boldmath $y$}|\mbox{\boldmath $x$}_{(\tilde{\mbox{\scriptsize \boldmath $w$}}, \tilde{\mbox{\scriptsize \boldmath $r$}})}\},\mbox{ for all } (\tilde{\mbox{\boldmath $w$}},\tilde{\mbox{\boldmath $r$}})\neq (\mbox{\boldmath $w$}, \mbox{\boldmath $r$}) , \mbox{\boldmath $r$},\tilde{\mbox{\boldmath $r$}} \in \mathcal{R},\nonumber\\
&&\mbox{C2: }-\frac{1}{N}\log Pr\{\mbox{\boldmath $y$}|\mbox{\boldmath $x$}_{(\mbox{\scriptsize \boldmath $w$}, \mbox{\scriptsize \boldmath $r$})}\} < \tau_{\mbox{\scriptsize \boldmath $r$}}(\mbox{\boldmath $y$}),
\label{CriteriaMulti}
\end{eqnarray}
where $\tau_{\mbox{\scriptsize \boldmath $r$}}(\cdot)$ is a pre-determined typicality threshold function of the channel output $\mbox{\boldmath $y$}$, associated with codewords of rate $\mbox{\boldmath $r$}$. If there is no codeword satisfying (\ref{CriteriaMulti}), the receiver reports a collision.

Define the corresponding exponent as the system error exponent $E_s=\lim_{N\to \infty} -\frac{1}{N} \log P_{es}$. Theorem \ref{Theorem2} implies the following achievable bound on $E_s$.
\begin{corollary}{\label{Corollary2}}
The system error exponent of single-user random access communication given in Theorem \ref{Theorem2} is lower-bounded by
\begin{equation}
E_s\ge \min\left\{ \min_{\mathcal{S}\subset\{1, \cdots, K\}}\min_{\mbox{\scriptsize \boldmath $r$}, \tilde{\mbox{\scriptsize \boldmath $r$}} \in \mathcal{R}, \mbox{\scriptsize \boldmath $r$}_{\mathcal{S}}= \tilde{\mbox{\scriptsize \boldmath $r$}}_{\mathcal{S}} }E_m(\mathcal{S}, \tilde{\mbox{\boldmath $r$}}, \mbox{\boldmath $P$}_{\mbox{\scriptsize \boldmath $X$}|\mbox{\scriptsize \boldmath $r$}}, \mbox{\boldmath $P$}_{\mbox{\scriptsize \boldmath $X$}|\tilde{\mbox{\scriptsize \boldmath $r$}}}),   \min_{\mathcal{S}\subset\{1, \cdots, K\}}\min_{\mbox{\scriptsize \boldmath $r$}\in \mathcal{R}, \tilde{\mbox{\scriptsize \boldmath $r$}} \not\in \mathcal{R}, \mbox{\scriptsize \boldmath $r$}_{\mathcal{S}}= \tilde{\mbox{\scriptsize \boldmath $r$}}_{\mathcal{S}}  }  E_i(\mathcal{S}, \mbox{\boldmath $r$}, \mbox{\boldmath $P$}_{\mbox{\scriptsize \boldmath $X$}|\mbox{\scriptsize \boldmath $r$}}, \mbox{\boldmath $P$}_{\mbox{\scriptsize \boldmath $X$}|\tilde{\mbox{\scriptsize \boldmath $r$}}})   \right\},
\label{MSysErrorExponentBound}
\end{equation}
where $E_m(\mathcal{S}, \tilde{\mbox{\boldmath $r$}}, \mbox{\boldmath $P$}_{\mbox{\scriptsize \boldmath $X$}|\mbox{\scriptsize \boldmath $r$}}, \mbox{\boldmath $P$}_{\mbox{\scriptsize \boldmath $X$}|\tilde{\mbox{\scriptsize \boldmath $r$}}})$ and $E_i(\mathcal{S}, \mbox{\boldmath $r$}, \mbox{\boldmath $P$}_{\mbox{\scriptsize \boldmath $X$}|\mbox{\scriptsize \boldmath $r$}}, \mbox{\boldmath $P$}_{\mbox{\scriptsize \boldmath $X$}|\tilde{\mbox{\scriptsize \boldmath $r$}}})$ are defined in (\ref{EmEiMulti}). $\QED$
\end{corollary}

Corollary \ref{Corollary2} is implied by Theorem \ref{Theorem2}.

As in the single-user system, if we define the decoding error exponent $E_d$ and the collision miss detection exponent $E_c$ as
\begin{eqnarray}
&& E_d= \min_{(\mbox{\scriptsize \boldmath $w$}, \mbox{\scriptsize \boldmath $r$}), \mbox{\scriptsize \boldmath $r$}\in \mathcal{R}, } \lim_{N\to \infty} -\frac{1}{N} \log P_e(\mbox{\boldmath $w$}, \mbox{\boldmath $r$} ), \nonumber \\
&& E_c= \min_{(\mbox{\scriptsize \boldmath $w$}, \mbox{\scriptsize \boldmath $r$}), \mbox{\scriptsize \boldmath $r$} \not\in \mathcal{R}} \lim_{N\to \infty} -\frac{1}{N} \log \bar{P}_c(\mbox{\boldmath $w$}, \mbox{\boldmath $r$} ),
\end{eqnarray}
then the system error exponent equals the minimum of the two exponents, i.e., $E_s=\min\{E_d, E_c\}$. Again, instead of optimizing the typicality function $\tau_{\mbox{\scriptsize \boldmath $r$}}(\cdot)$ to lower bound $E_s$, $\tau_{\mbox{\scriptsize \boldmath $r$}}(\cdot)$ can be used to adjust the tradeoff between $E_d$ and $E_c$.

Note that, in Theorem \ref{Theorem2}, the receiver either decodes the messages of {\it all} users or reports a collision for {\it all} users. In practice, the receiver could choose to output message estimates for a subset of users and to report collision for the others. The corresponding achievable communication rate region has been given in \cite{ref Luo09}. An error performance bound can be derived using an approach similar to the one shown in the proof of Theorem \ref{Theorem2}. The detailed analysis, however, is skipped.

\section{Error Performance under Generalized Random Coding with Standard Communication Rate}
\label{Section StandardRate}

In the previous sections, we used the practical definition of communication rate, i.e., communication rate equals the normalized data nats per symbol encoded in a packet. Codewords of each user are partitioned into $M$ classes each corresponding to a rate option. This is equivalent to indexing the codewords using a message and rate pair $(w, r)$. We assumed codeword symbols within each class, i.e., corresponding to the same $r$, should be randomly generated according to the same input distribution. In this section, we extend the results to the generalized random coding scheme \cite{ref Luo09} where symbols of different codewords, as opposed to different codeword classes, can be generated according to different input distributions.

We will index the codewords in a codebook using a macro message $W$, which is essentially another expression of the $(w, r)$ pair used in previous sections. In other words, $W$ contains both information about the message $w$ and the rate $r$ in practical senses. The generalized random coding scheme is defined originally in \cite{ref Luo09} as follows.

\begin{definition}{\label{Definition1}} ({\bf generalized random coding} \cite{ref Luo09})
Let ${\cal L}=\{{\cal C}_{\theta}: \theta \in \Theta\}$ be a library of codebooks. Each codebook in the library contains $e^{NR_{\max}}$ codewords of length $N$, where $R_{\max}$ is an arbitrary large finite constant. Let the codebooks be indexed by a set $\Theta$. Let the actual codebook chosen by the transmitter be ${\cal C}_{\theta}$ where the index $\theta$ is a random variable following distribution $\gamma$. Let $W\in \{1, \cdots, e^{NR_{\max}} \}$ be a macro message used to index the codewords in each codebook. Denote ${\cal C}_{\theta}(W)_j$ as the $j^{th}$ symbol of the codeword corresponding to macro message $W$ in codebook ${\cal C}_{\theta}$. We define $({\cal L}, \gamma)$ as a {\it generalized random coding scheme} following distribution $P_{X|W}$, if the random variables $X_{W,j} : \theta \to {\cal C}_{\theta}(W)_j$, $\forall j, W$, are independently distributed according to input distribution $P_{X|W}$. $\QED$
\end{definition}

Note that a generalized random coding scheme allows codeword symbols corresponding to different messages to be generated according to different input distributions. Because codewords are indexed using macro message $W$, communication rate $r$ becomes a function of $W$. Consequently, the practical communication rate $r$ used in previous sections only represents a specific choice of the rate function. In order to distinguish codewords from each other in rate and error performance characterization, in this section, we will switch to the following {\it standard communication rate} definition, originally introduced in \cite{ref Luo09}.

\begin{definition}{\label{Definition2}} ({\bf standard communication rate} \cite{ref Luo09})
Assume codebook ${\cal C}$ has $e^{NR_{\max}}$ codewords of length $N$, where $R_{\max}$ is an arbitrary large finite constant. Let the corresponding messages or codewords be indexed by $W\in \{1, \cdots, e^{NR_{\max}}\}$. For each message $W$, we define its {\it standard communication rate}, in nats per symbol, as $r(W)=\frac{1}{N}\log W$. $\QED$
\end{definition}

Since the standard rate function $r(W)=\frac{1}{N}\log W$ is invertible, system performance characterized in any other rate function can be derived from that of the standard rate function \cite{ref Luo09}\footnote{Note that the standard rate is defined using the natural log in this paper, while it was defined using the base-2 log in \cite{ref Luo09}.}.

The following definition specifies a sequence of generalized random coding schemes following an asymptotic input distribution.

\begin{definition}{\label{Definition3}} ({\bf asymptotic input distribution} \cite{ref Luo09})
Let $\{({\cal L}^{(N)}, \gamma^{(N)} )\}$ be a sequence of random coding schemes, where $({\cal L}^{(N)}, \gamma^{(N)} )$ is a generalized random coding scheme with codeword length $N$ and input distribution $P^{(N)}_{X|W^{(N)}}$. Assume each codebook in library ${\cal L}^{(N)}$ has $e^{NR_{\max}}$ codewords. Let $P_{X|r}$ be an input distribution defined as a function of the standard rate $r$, for all $r\in [0, R_{\max}]$. We say $\{({\cal L}^{(N)}, \gamma^{(N)} )\}$ follows an asymptotic input distribution $P_{X|r}$, if for all $\{W^{(N)}\}$ sequences with well defined rate limit $\lim_{N\to\infty}r(W^{(N)})$, we have
\begin{equation}
\lim_{N\to \infty}P_{X|W^{(N)}}^{(N)}=\lim_{N\to \infty} P_{X| r(W^{(N)})}.
\end{equation}
Note that since we do not assume $P_{X|r}$ is continuous in $r$, we may not have $\lim_{N\to \infty} P_{X| r(W^{(N)})}=P_{X|\lim_{N\to\infty}r(W^{(N)})}$. $\QED$
\end{definition}

Let us still use bold font vector variables to denote the corresponding variables of multiple users. Theorem \ref{Theorem3} gives the achievable error exponent of a random multiple access system using generalized random coding.

\begin{theorem}{\label{Theorem3}}
Consider $K$-user random multiple access communication over a discrete-time memoryless channel $P_{Y|\mbox{\scriptsize \boldmath  $X$}}$ using a sequence of generalized random coding schemes $\{(\mbox{\boldmath  ${\cal L}$}^{(N)}, \mbox{\boldmath  $\gamma$}^{(N)})\}$. Assume $\{(\mbox{\boldmath  ${\cal L}$}^{(N)}, \mbox{\boldmath  $\gamma$}^{(N)})\}$ follows asymptotic distribution $\mbox{\boldmath  $P$}_{\mbox{\scriptsize \boldmath  $X$}|\mbox{\scriptsize \boldmath  $r$}}$. For any user $k$, assume $P_{X_k|r_k}$ is only discontinuous in $r_k$ at a finite number of points. Let the operation region $\mathcal{R}$ be strictly contained in an achievable rate region, specified in \cite{ref Luo09}. Equation (\ref{MSysErrorExponentBound}) gives an achievable lower bound on the system error exponent $E_s$, with rates in the equation being the standard communication rates. $\QED$
\end{theorem}

The proof of Theorem \ref{Theorem3} is given in Appendix \ref{ProofTheorem3}. In the proof, an achievable error probability bound in the case of a finite codeword length is also given in Lemma \ref{Lemma1}.

\section{Conclusions}
\label{Section Conclusion}

We investigated the error performance of a new coding scheme for random access communication over discrete-time memoryless channels. Two types of error events are considered, the decoding error event when the transmitted communication rate vector is inside the operation region, and the collision miss detection event when the transmitted communication rate vector is outside the operation region. Upper bound on the system error probability, defined as the maximum probability of both error events, is derived for both single-user random access and random multiple access communication systems with a finite codeword length. We showed that, if the operation region is strictly contained in an achievable rate region, then the system error probability can decrease exponentially in the codeword length. An achievable lower bound on the system error exponent is obtained. The result is also extended to random multiple access communication systems using generalized random coding with standard communication rate definition.

\appendix

\subsection{Proof of Theorem \ref{Theorem1}}
\label{ProofTheorem1}
\begin{proof}
To derive the system error probability upper bound, we assume the receiver uses the decoding algorithm whose decoding criteria are specified in (\ref{CriteriaSingle}).

We next define three probability terms that will be extensively used in the probability bound derivation.

Fist, assume $(w, r)$ is the transmitted message and rate pair with $r\in \mathcal{R}$. We define $P_{m[r, \tilde{r}]}$ as the probability that the receiver finds another codeword with rate $\tilde{r} \in \mathcal{R}$ that has a likelihood value no worse than the transmitted codeword.
\begin{equation}
P_{m[r, \tilde{r}]}= Pr\left\{ P(\mbox{\boldmath $y$}| \mbox{\boldmath $x$}_{(w, r)}) \le  P(\mbox{\boldmath $y$}| \mbox{\boldmath $x$}_{(\tilde{w}, \tilde{r})})\right\}, \qquad (\tilde{w}, \tilde{r})\ne (w, r), \tilde{r}\in \mathcal{R}.
\end{equation}

Second, assume $(w, r)$ is the transmitted message and rate pair with $r\in \mathcal{R}$. We define $P_{tr}$ as the probability that the likelihood of the transmitted codeword is below a predetermined threshold.
\begin{equation}
P_{tr}= Pr \left \{ P(\mbox{\boldmath $y$}| \mbox{\boldmath $x$}_{(w, r)}) \le  e^{-N\tau_r(\mbox{\scriptsize \boldmath $y$})}\right\},
\end{equation}
where $\tau_r(\mbox{\boldmath $y$})$ is a threshold, as a function of $r$ and $\mbox{\boldmath $y$}$, that will be optimized later\footnote{Note that the subscript $r$ of $\tau_r(\mbox{\boldmath $y$})$ represents the corresponding estimated rate of the receiver output. Although with an abuse of the notation, we occasionally use the same symbol $r$ to denote both the transmitted rate and the corresponding rate estimation at the receiver, it is important to note that we do not assume the receiver should know the transmitted rate.}.

Third, assume $(\tilde{w}, \tilde{r})$ is the transmitted message and rate pair with $\tilde{r} \not\in \mathcal{R}$. We define $P_{i[\tilde{r}, r]}$ as the probability that the receiver finds another codeword with rate $r\in \mathcal{R}$ that has a likelihood value above the required threshold.
\begin{equation}
P_{i[\tilde{r}, r]}= Pr\left\{ P(\mbox{\boldmath $y$}| \mbox{\boldmath $x$}_{(w, r)}) > e^{-N\tau_{r}(\mbox{\scriptsize \boldmath $y$})}\right\}, \qquad (w, r)\ne (\tilde{w}, \tilde{r}), r\in \mathcal{R}.
\end{equation}

With these probability definitions, we can upper bound the system error probability $P_{es}$ by
\begin{equation}
P_{es} \le \max \left\{ \max_{r \in \mathcal{R}} \sum_{\tilde{r}\in \mathcal{R}} P_{m[r, \tilde{r}]}+P_{tr}, \max_{\tilde{r} \not\in \mathcal{R}}\sum_{r\in R}P_{i[\tilde{r}, r]}  \right\}.
\label{PesBound1}
\end{equation}
Next, we will upper bound each of the probability terms on the right hand side of (\ref{PesBound1}).

{\bf Step 1: } Upper-bounding $P_{m[r, \tilde{r}]}$.

Assume $(w, r)$ is the transmitted message and rate pair with $r\in \mathcal{R}$. Given $r, \tilde{r} \in \mathcal{R}$, $P_{m[r, \tilde{r}]}$ can be written as
\begin{equation}
P_{m[r, \tilde{r}]}=E_{\theta}\left[\sum_{\mbox{\scriptsize \boldmath $y$} }P(\mbox{\boldmath $y$}| \mbox{\boldmath $x$}_{(w, r)}) \phi_{m[r, \tilde{r}]}( \mbox{\boldmath $y$} ) \right],
\end{equation}
where $\phi_{m[r, \tilde{r}]}( \mbox{\boldmath $y$} )=1$ if $P(\mbox{\boldmath $y$}| \mbox{\boldmath $x$}_{(w, r)}) \le P(\mbox{\boldmath $y$}| \mbox{\boldmath $x$}_{(\tilde{w}, \tilde{r})})$ for some $(\tilde{w}, \tilde{r}) \ne (w, r)$, and  $\phi_{m[r, \tilde{r}]}( \mbox{\boldmath $y$} )=0$ otherwise.

Revised from Gallager's approach \cite{ref Gallager65}, for any $\rho>0$ and $s>0$, we can bound $\phi_{m[r, \tilde{r}]}( \mbox{\boldmath $y$} )$ by
\begin{equation}
\phi_{m[r, \tilde{r}]}( \mbox{\boldmath $y$} )\le \left[ \frac{\sum_{\tilde{w}, (\tilde{w}, \tilde{r})\ne (w, r)}P(\mbox{\boldmath $y$}| \mbox{\boldmath $x$}_{(\tilde{w}, \tilde{r})})^{\frac{s}{\rho}} }{P(\mbox{\boldmath $y$}| \mbox{\boldmath $x$}_{(w, r)})^{\frac{s}{\rho}}} \right]^{\rho}, \quad \rho>0, s>0.
\end{equation}
Consequently, $P_{m[r, \tilde{r}]}$ is upper bounded by
\begin{eqnarray}
P_{m[r, \tilde{r}]} &\le & E_{\theta}\left[\sum_{\mbox{\scriptsize \boldmath $y$} } P(\mbox{\boldmath $y$}| \mbox{\boldmath $x$}_{(w, r)}) \left[ \frac{\sum_{\tilde{w}, (\tilde{w}, \tilde{r})\ne (w, r)}P(\mbox{\boldmath $y$}| \mbox{\boldmath $x$}_{(\tilde{w}, \tilde{r})})^{\frac{s}{\rho}} }{P(\mbox{\boldmath $y$}| \mbox{\boldmath $x$}_{(w, r)})^{\frac{s}{\rho}}} \right]^{\rho} \right] \nonumber \\
&=& E_{\theta}\left[\sum_{\mbox{\scriptsize \boldmath $y$} }  P(\mbox{\boldmath $y$}| \mbox{\boldmath $x$}_{(w, r)})^{1-s} \left[ \sum_{\tilde{w}, (\tilde{w}, \tilde{r})\ne (w, r)}P(\mbox{\boldmath $y$}| \mbox{\boldmath $x$}_{(\tilde{w}, \tilde{r})})^{\frac{s}{\rho}} \right]^{\rho}  \right] \nonumber \\
&=&\sum_{\mbox{\scriptsize \boldmath $y$} }   E_{\theta}\left[P(\mbox{\boldmath $y$}| \mbox{\boldmath $x$}_{(w, r)})^{1-s}\right]  E_{\theta}\left[\left[ \sum_{\tilde{w}, (\tilde{w}, \tilde{r})\ne (w, r)}P(\mbox{\boldmath $y$}| \mbox{\boldmath $x$}_{(\tilde{w}, \tilde{r})})^{\frac{s}{\rho}} \right]^{\rho}  \right],
\label{Inequality1.1}
\end{eqnarray}
where in the last step, we can separate the expectation operations due to independence between $\mbox{\boldmath $x$}_{(w, r)}$ and $\mbox{\boldmath $x$}_{(\tilde{w}, \tilde{r})}$.

Now assume $0<\rho\le 1$. Inequality (\ref{Inequality1.1}) can be further bounded by
\begin{eqnarray}
P_{m[r, \tilde{r}]} & \le & \sum_{\mbox{\scriptsize \boldmath $y$} }   E_{\theta}\left[P(\mbox{\boldmath $y$}| \mbox{\boldmath $x$}_{(w, r)})^{1-s}\right]  E_{\theta}\left[\left[ \sum_{\tilde{w}}P(\mbox{\boldmath $y$}| \mbox{\boldmath $x$}_{(\tilde{w}, \tilde{r})})^{\frac{s}{\rho}} \right]^{\rho}  \right] \nonumber \\
& \le & e^{N\rho\tilde{r}} \sum_{\mbox{\scriptsize \boldmath $y$} }   E_{\theta}\left[P(\mbox{\boldmath $y$}| \mbox{\boldmath $x$}_{(w, r)})^{1-s}\right]  \left[  E_{\theta}\left[P(\mbox{\boldmath $y$}| \mbox{\boldmath $x$}_{(\tilde{w}, \tilde{r})})^{\frac{s}{\rho}} \right] \right]^{\rho} \nonumber \\
&=& e^{N\rho\tilde{r}} \left\{ \sum_Y\left[\sum_X P_{X|r}(X)P(Y|X)^{1-s} \right]\left[ \sum_X P_{X|\tilde{r}}(X) P(Y|X)^{\frac{s}{\rho}}\right]^\rho     \right\}^N .
\label{Inequality1.2}
\end{eqnarray}

Since (\ref{Inequality1.2}) holds for all $0< \rho\le 1$, $s>0$, and it is easy to verify that the bound becomes trivial for $s>1$, we have
\begin{equation}
P_{m[r, \tilde{r}]} \le \exp \left\{-NE_m(\tilde{r}, P_{X|r}, P_{X|\tilde{r}}) \right\},
\label{ProofPmBound}
\end{equation}
where $E_m(\tilde{r}, P_{X|r}, P_{X|\tilde{r}})$ is given by
\begin{equation}
E_m(\tilde{r}, P_{X|r}, P_{X|\tilde{r}})=\max_{0<\rho\le 1} -\rho \tilde{r}+\max_{0<s\le 1} -\log \sum_Y\left[\sum_X P_{X|r}(X)P(Y|X)^{1-s} \right]\left[ \sum_X P_{X|\tilde{r}}(X) P(Y|X)^{\frac{s}{\rho}}\right]^\rho.
\label{ProofEmBound}
\end{equation}

{\bf Step 2: } Upper-bounding $P_{tr}$.

Assume $(w, r)$ is the transmitted message and rate pair with $r\in \mathcal{R}$. Rewrite $P_{tr}$ as
\begin{equation}
P_{tr}=E_{\theta}\left[\sum_{\mbox{\scriptsize \boldmath $y$} }P(\mbox{\boldmath $y$}| \mbox{\boldmath $x$}_{(w, r)})\phi_{tr}(\mbox{\boldmath $y$})  \right],
\end{equation}
where $\phi_{tr}(\mbox{\boldmath $y$})=1$ if $P(\mbox{\boldmath $y$}| \mbox{\boldmath $x$}_{(w, r)})\le e^{-N\tau_r(\mbox{\boldmath $y$})}$, otherwise $\phi_{tr}(\mbox{\boldmath $y$})=0$. Note that the value of $\tau_r(\mbox{\boldmath $y$})$ will be specified later.

For any $s_1>0$, we can bound $\phi_{tr}(\mbox{\boldmath $y$})$ as
\begin{equation}
\phi_{tr}(\mbox{\boldmath $y$})\le \frac{e^{-Ns_1 \tau_r(\mbox{\scriptsize \boldmath $y$}) }}{ P(\mbox{\boldmath $y$}| \mbox{\boldmath $x$}_{(w, r)})^{s_1}} , \quad s_1>0.
\end{equation}
This yields
\begin{eqnarray}
P_{tr} &\le & E_{\theta}\left[ \sum_{\mbox{\scriptsize \boldmath $y$} }P(\mbox{\boldmath $y$}| \mbox{\boldmath $x$}_{(w, r)})^{1-s_1} e^{-Ns_1 \tau_r(\mbox{\scriptsize \boldmath $y$}) } \right] \nonumber \\
&=& \sum_{\mbox{\scriptsize \boldmath $y$} } E_{\theta}\left[ P(\mbox{\boldmath $y$}| \mbox{\boldmath $x$}_{(w, r)})^{1-s_1}\right] e^{-Ns_1 \tau_r(\mbox{\scriptsize \boldmath $y$}) }.
\label{Bound2}
\end{eqnarray}
We will come back to this inequality later when we optimize $\tau_r(\mbox{ \boldmath $y$})$.

{\bf Step 3: } Upper-bounding $P_{i[\tilde{r}, r]}$.

Assume $(\tilde{w}, \tilde{r})$ is the transmitted message and rate pair with $\tilde{r} \not\in \mathcal{R}$. Given $r \in \mathcal{R}$, we first rewrite $P_{i[\tilde{r}, r]}$ as
\begin{equation}
P_{i[\tilde{r}, r]}=E_{\theta}\left[ \sum_{\mbox{\scriptsize \boldmath $y$} }P(\mbox{\boldmath $y$}| \mbox{\boldmath $x$}_{(\tilde{w}, \tilde{r})})\phi_{i[\tilde{r}, r]}(\mbox{\boldmath $y$} ) \right],
\end{equation}
where $\phi_{i[\tilde{r}, r]}(\mbox{\boldmath $y$} )=1$ if there exists $(w, r)$ with $r \in \mathcal{R}$ to satisfy $P(\mbox{\boldmath $y$}| \mbox{\boldmath $x$}_{(w, r)})> e^{-N\tau_r(\mbox{\scriptsize \boldmath $y$} )}$, otherwise $\phi_{i[\tilde{r}, r]}(\mbox{\boldmath $y$} )=0$.

For any $s_2>0$ and $\tilde{\rho}>0$, we can bound $\phi_{i[\tilde{r}, r]}(\mbox{\boldmath $y$} )$ by
\begin{equation}
\phi_{i[\tilde{r}, r]}(\mbox{\boldmath $y$} ) \le \left[ \frac{\sum_{w}P(\mbox{\boldmath $y$}| \mbox{\boldmath $x$}_{(w, r)})^{\frac{s_2}{\tilde{\rho}}}}{e^{-N\frac{s_2}{\tilde{\rho}}\tau_r(\mbox{\scriptsize \boldmath $y$} )}}  \right]^{\tilde{\rho}}, \quad s_2>0, \tilde{\rho} >0.
\end{equation}
This gives,
\begin{eqnarray}
P_{i[\tilde{r}, r]} &\le & E_{\theta}\left[ \sum_{\mbox{\scriptsize \boldmath $y$} }P(\mbox{\boldmath $y$}| \mbox{\boldmath $x$}_{(\tilde{w}, \tilde{r})})\left[ \sum_wP(\mbox{\boldmath $y$}| \mbox{\boldmath $x$}_{(w, r)})^{\frac{s_2}{\tilde{\rho}}}  \right]^{\tilde{\rho}} e^{Ns_2\tau_r(\mbox{\scriptsize \boldmath $y$} )} \right] \nonumber \\
& = & \sum_{\mbox{\scriptsize \boldmath $y$} } E_{\theta}\left[P(\mbox{\boldmath $y$}| \mbox{\boldmath $x$}_{(\tilde{w}, \tilde{r})}) \right]E_{\theta}\left[ \left[\sum_w P(\mbox{\boldmath $y$}| \mbox{\boldmath $x$}_{(w, r)})^{\frac{s_2}{\tilde{\rho}}} \right]^{\tilde{\rho}} \right]e^{Ns_2\tau_r(\mbox{\scriptsize \boldmath $y$} )} .
\label{Inequality1.3}
\end{eqnarray}
Note that we can separate the expectation operators in the last step due to independence between $\mbox{\boldmath $x$}_{(w, r)}$ and $\mbox{\boldmath $x$}_{(\tilde{w}, \tilde{r})}$.

Assume $0<\tilde{\rho}\le 1$. Inequality (\ref{Inequality1.3}) leads to
\begin{eqnarray}
P_{i[\tilde{r}, r]} &\le& \sum_{\mbox{\scriptsize \boldmath $y$} } E_{\theta}\left[P(\mbox{\boldmath $y$}| \mbox{\boldmath $x$}_{(\tilde{w}, \tilde{r})}) \right]\left[E_{\theta}\left[P(\mbox{\boldmath $y$}| \mbox{\boldmath $x$}_{(w, r)})^{\frac{s_2}{\tilde{\rho}}} \right] \right]^{\tilde{\rho}} e^{Ns_2\tau_r(\mbox{\scriptsize \boldmath $y$} )} e^{N\tilde{\rho}r} \nonumber \\
& \le & \max_{\tilde{r}\not \in \mathcal{R}} \sum_{\mbox{\scriptsize \boldmath $y$} } E_{\theta}\left[P(\mbox{\boldmath $y$}| \mbox{\boldmath $x$}_{(\tilde{w}, \tilde{r})}) \right]\left\{E_{\theta}\left[P(\mbox{\boldmath $y$}| \mbox{\boldmath $x$}_{(w, r)})^{\frac{s_2}{\tilde{\rho}}} \right] \right\}^{\tilde{\rho}} e^{Ns_2\tau_r(\mbox{\scriptsize \boldmath $y$} )} e^{N\tilde{\rho}r}.
\label{Bound3}
\end{eqnarray}
Note that the bound obtained in the last step is no longer a function of $\tilde{r}$.

{\bf Step 4: } Choosing $\tau_{r}(\mbox{\boldmath $y$} )$.

In this step, we determine the typicality threshold $\tau_{r}(\mbox{\boldmath $y$} )$ that optimizes the bounds in (\ref{Bound2}) and (\ref{Bound3}).

Let us define $\tilde{r}^*\not \in \mathcal{R}$ as
\begin{equation}
\tilde{r}^*=\mathop{\mbox{argmax}}_{\tilde{r}\not \in \mathcal{R}} \sum_{\mbox{\scriptsize \boldmath $y$} } E_{\theta}\left[P(\mbox{\boldmath $y$}| \mbox{\boldmath $x$}_{(\tilde{w}, \tilde{r})}) \right]\left\{E_{\theta}\left[P(\mbox{\boldmath $y$}| \mbox{\boldmath $x$}_{(w, r)})^{\frac{s_2}{\tilde{\rho}}} \right] \right\}^{\tilde{\rho}} e^{Ns_2\tau_r(\mbox{\scriptsize \boldmath $y$} )} e^{N\tilde{\rho}r}.
\end{equation}
Given $r\in \mathcal{R}$, $\mbox{\boldmath $y$}$, and the auxiliary variables $s_1>0$, $s_2>0$, $0<\tilde{\rho}\le 1$, we choose $\tau_{r}(\mbox{\boldmath $y$} )$ such that the following equality holds,
\begin{equation}
E_{\theta}\left[ P(\mbox{\boldmath $y$}| \mbox{\boldmath $x$}_{(w, r)})^{1-s_1}\right] e^{-Ns_1 \tau_r(\mbox{\scriptsize \boldmath $y$}) } = E_{\theta}\left[P(\mbox{\boldmath $y$}| \mbox{\boldmath $x$}_{(\tilde{w}, \tilde{r}^*)}) \right]\left\{E_{\theta}\left[P(\mbox{\boldmath $y$}| \mbox{\boldmath $x$}_{(w, r)})^{\frac{s_2}{\tilde{\rho}}} \right] \right\}^{\tilde{\rho}} e^{Ns_2\tau_r(\mbox{\scriptsize \boldmath $y$} )} e^{N\tilde{\rho}r}.
\label{Inequality1.5}
\end{equation}
This is always possible since the left hand side of (\ref{Inequality1.5}) decreases in $\tau_{r}(\mbox{\boldmath $y$} )$ while the right hand side of (\ref{Inequality1.5}) increases in $\tau_{r}(\mbox{\boldmath $y$} )$.

Equation (\ref{Inequality1.5}) implies
\begin{equation}
e^{-N \tau_r(\mbox{\scriptsize \boldmath $y$}) } =\frac{ \left\{E_{\theta}\left[P(\mbox{\boldmath $y$}| \mbox{\boldmath $x$}_{(\tilde{w}, \tilde{r}^*)}) \right]\right\}^{\frac{1}{s_1+s_2}}\left\{E_{\theta}\left[P(\mbox{\boldmath $y$}| \mbox{\boldmath $x$}_{(w, r)})^{\frac{s_2}{\tilde{\rho}}} \right] \right\}^{\frac{\tilde{\rho}}{s_1+s_2}} e^{N\frac{\tilde{\rho}}{s_1+s_2}r} } { \left\{E_{\theta}\left[ P(\mbox{\boldmath $y$}| \mbox{\boldmath $x$}_{(w, r)})^{1-s_1}\right]\right\}^{\frac{1}{s_1+s_2}} }.
\label{Equality1.6}
\end{equation}

Substituting (\ref{Equality1.6}) into (\ref{Bound2}) yields
\begin{equation}
P_{tr} \le  \sum_{\mbox{\scriptsize \boldmath $y$} } \left\{E_{\theta}\left[ P(\mbox{\boldmath $y$}| \mbox{\boldmath $x$}_{(w, r)})^{1-s_1}\right]\right\}^{\frac{s_2}{s_1+s_2}} \left\{E_{\theta}\left[P(\mbox{\boldmath $y$}| \mbox{\boldmath $x$}_{(\tilde{w}, \tilde{r}^*)}) \right]\right\}^{\frac{s_1}{s_1+s_2}}\left\{E_{\theta}\left[P(\mbox{\boldmath $y$}| \mbox{\boldmath $x$}_{(w, r)})^{\frac{s_2}{\tilde{\rho}}} \right] \right\}^{\frac{s_1\tilde{\rho}}{s_1+s_2}} e^{N\frac{s_1\tilde{\rho}}{s_1+s_2}r} .
\label{Equality1.7}
\end{equation}

Let $s_2 < \tilde{\rho}$ and $s_1=1-\frac{s_2}{\tilde{\rho}}$. Inequality (\ref{Equality1.7}) becomes
\begin{equation}
P_{tr} \le \sum_{\mbox{\scriptsize \boldmath $y$} } \left\{E_{\theta}\left[ P(\mbox{\boldmath $y$}| \mbox{\boldmath $x$}_{(w, r)})^{\frac{s_2}{\tilde{\rho}}}\right]\right\}^{\frac{ \tilde{\rho}^2}{\tilde{\rho}-(1-\tilde{\rho})s_2}} \left\{ E_{\theta}\left[P(\mbox{\boldmath $y$}| \mbox{\boldmath $x$}_{(\tilde{w}, \tilde{r}^*)}) \right]\right\}^{\frac{\tilde{\rho}-s_2}{\tilde{\rho}-(1-\tilde{\rho})s_2}}e^{N\frac{\tilde{\rho}(\tilde{\rho}-s_2)}{\tilde{\rho}-(1-\tilde{\rho})s_2}r}.
\label{Equality1.8}
\end{equation}

Now do a variable change with $\rho=\frac{\tilde{\rho}(\tilde{\rho}-s_2)}{\tilde{\rho}-(1-\tilde{\rho})s_2}$ and $s=1-\frac{\tilde{\rho}-s_2}{\tilde{\rho}-(1-\tilde{\rho})s_2}$, and note that $s+\rho\le 1$. Inequality (\ref{Equality1.8}) becomes
\begin{eqnarray}
P_{tr} & \le & \sum_{\mbox{\scriptsize \boldmath $y$} } \left\{E_{\theta}\left[ P(\mbox{\boldmath $y$}| \mbox{\boldmath $x$}_{(w, r)})^{\frac{s}{s+\rho}}\right]\right\}^{s+\rho} \left\{E_{\theta}\left[P(\mbox{\boldmath $y$}| \mbox{\boldmath $x$}_{(\tilde{w}, \tilde{r}^*)}) \right]\right\}^{1-s}e^{N\rho r} \nonumber \\
&\le & \max_{\tilde{r}\not\in \mathcal{R}} \left\{ \sum_Y \left[ \sum_X P_{X|r}(X)P(Y|X)^{\frac{s}{s+\rho}}\right]^{s+\rho}\left[\sum_X P_{X|\tilde{r}}(X)P(Y|X) \right]^{1-s} \right\}^Ne^{N\rho r}.
\label{Equality1.9}
\end{eqnarray}

Following the same derivation, we can see that $P_{i[\tilde{r}, r]}$ is also upper-bounded by the right hand side of (\ref{Equality1.9}). Because (\ref{Equality1.9}) holds for all $0<\rho \le 1$ and $0<s \le 1-\rho$, we have
\begin{equation}
P_{tr}, P_{i[\tilde{r}, r]} \le \max_{\tilde{r} \not\in \mathcal{R}} \exp\{ -N E_i(r, P_{X|r}, P_{X|\tilde{r}})\},
\label{ProofPiBound}
\end{equation}
where
\begin{equation}
E_i(r, P_{X|r}, P_{X|\tilde{r}})=\max_{0<\rho\le 1} -\rho r+\max_{0<s\le 1-\rho} -\log \sum_Y\left[ \sum_X P_{X|r}(X)P(Y|X)^{\frac{s}{s+\rho}}\right]^{s+\rho}\left[\sum_X P_{X|\tilde{r}}(X)P(Y|X)\right]^{1-s}.
\label{ProofEiBound}
\end{equation}

Finally, substituting (\ref{ProofPmBound}) and (\ref{ProofPiBound}) into (\ref{PesBound1}) gives the desired result.

\end{proof}

\subsection{Proof of Theorem \ref{Theorem2}}
\label{ProofTheorem2}
\begin{proof}
Due to the involvement of multiple users, notations used in this proof are rather complicated. To make the proof easy to follow, we carefully organize the derivations according to the same structure as the proof of Theorem \ref{Theorem1}. Because Theorem \ref{Theorem1} is indeed a simplified single-user version of Theorem \ref{Theorem2}, it will help significantly if the reader follows the proof of Theorem \ref{Theorem2} by comparing it, step by step, to the proof of Theorem \ref{Theorem1}.

We assume the receiver uses the decoding algorithm whose decoding criteria are specified in (\ref{CriteriaMulti}). However, to facilitate the derivation, we first need to make a minor revision to the decoding rules.

Given the received channel symbols $\mbox{\boldmath $y$}$, the receiver outputs a message and rate vector pair $(\mbox{\boldmath $w$}, \mbox{\boldmath $r$})$, with $\mbox{\boldmath $r$}\in \mathcal{R}$, if for {\it all} user subsets $\mathcal{S}\subset \{1, \cdots, K\}$, the following two conditions are met.
\begin{eqnarray}
&&\mbox{C1R: }-\frac{1}{N}\log Pr\{\mbox{\boldmath $y$}|\mbox{\boldmath $x$}_{(\mbox{\scriptsize \boldmath $w$}, \mbox{\scriptsize \boldmath $r$})}\} < -\frac{1}{N}\log Pr\{\mbox{\boldmath $y$}|\mbox{\boldmath $x$}_{(\tilde{\mbox{\scriptsize \boldmath $w$}}, \tilde{\mbox{\scriptsize \boldmath $r$}})}\}, \nonumber \\
&& \qquad \qquad \mbox{ for all } (\tilde{\mbox{\boldmath $w$}},\tilde{\mbox{\boldmath $r$}}) \mbox{ with } \tilde{\mbox{\boldmath $r$}} \in \mathcal{R}, (\tilde{\mbox{\boldmath $w$}}_{\mathcal{S}}, \tilde{\mbox{\boldmath $r$}}_{\mathcal{S}})=(\mbox{\boldmath $w$}_{\mathcal{S}}, \mbox{\boldmath $r$}_{\mathcal{S}}), \mbox{ and } (\tilde{w}_k, \tilde{r}_k)\neq (w_k, r_k), \forall k\not\in \mathcal{S}, \nonumber\\
&&\mbox{C2R: }-\frac{1}{N}\log Pr\{\mbox{\boldmath $y$}|\mbox{\boldmath $x$}_{(\mbox{\scriptsize \boldmath $w$}, \mbox{\scriptsize \boldmath $r$})}\} < \tau_{(\mbox{\scriptsize \boldmath $r$}, \mathcal{S})}(\mbox{\boldmath $y$}).
\label{CriteriaMultiRevised}
\end{eqnarray}
Note that in Condition C1R, we added the requirements of $(\tilde{\mbox{\boldmath $w$}}_{\mathcal{S}}, \tilde{\mbox{\boldmath $r$}}_{\mathcal{S}})=(\mbox{\boldmath $w$}_{\mathcal{S}}, \mbox{\boldmath $r$}_{\mathcal{S}})$ and $(\tilde{w}_k, \tilde{r}_k)\neq (w_k, r_k)$, $\forall k\not\in \mathcal{S}$. The union of Conditions C1R over all user subsets $\mathcal{S}\subset \{1, \cdots, K\}$ gives Condition C1 in (\ref{CriteriaMulti}). In Condition C2R, we assume the typicality threshold $\tau_{(\mbox{\scriptsize \boldmath $r$}, \mathcal{S})}(\mbox{\boldmath $y$})$ depends on both $\mbox{\boldmath $r$}$ and $\mathcal{S}$. By taking the union over $\mathcal{S}\subset \{1, \cdots, K\}$, Condition C2R in (\ref{CriteriaMultiRevised}) implies that the typicality threshold in Condition C2 of (\ref{CriteriaMulti}) should be set at $\tau_{\mbox{\scriptsize \boldmath $r$}}(\mbox{\boldmath $y$})=\min_{\mathcal{S}\subset \{1, \cdots, K\}}\tau_{(\mbox{\scriptsize \boldmath $r$}, \mathcal{S})}(\mbox{\boldmath $y$})$. In the rest of the proof, we will analyze the probabilities and optimize the thresholds $\tau_{(\mbox{\scriptsize \boldmath $r$}, \mathcal{S})}(\mbox{\boldmath $y$})$ separately for different $\mathcal{S}$.

Given a user subset $\mathcal{S}\subset \{1, \cdots, K\}$, we define the following three probability terms that will be extensively used in the probability bound derivation.

First, assume $(\mbox{\boldmath $w$}, \mbox{\boldmath $r$})$ is the transmitted message and rate pair with $\mbox{\boldmath $r$}\in \mathcal{R}$. We define $P_{m[\mbox{\scriptsize \boldmath $r$}, \tilde{\mbox{\scriptsize \boldmath $r$}}, \mathcal{S}]}$ as the probability that the receiver finds another message and rate pair $(\tilde{\mbox{\boldmath $w$}}, \tilde{\mbox{\boldmath $r$}})$ with $\tilde{\mbox{\boldmath $r$}} \in \mathcal{R}$, $(\tilde{\mbox{\boldmath $w$}}_{\mathcal{S}}, \tilde{\mbox{\boldmath $r$}}_{\mathcal{S}})=(\mbox{\boldmath $w$}_{\mathcal{S}}, \mbox{\boldmath $r$}_{\mathcal{S}})$, and  $(\tilde{w}_k, \tilde{r}_k)\neq (w_k, r_k), \forall k\not\in \mathcal{S}$, that has a likelihood value no worse than the transmitted codeword.
\begin{eqnarray}
&& P_{m[\mbox{\scriptsize \boldmath $r$}, \tilde{\mbox{\scriptsize \boldmath $r$}}, \mathcal{S}]}= Pr\left\{ P(\mbox{\boldmath $y$}| \mbox{\boldmath $x$}_{(\mbox{\scriptsize \boldmath $w$}, \mbox{\scriptsize \boldmath $r$})}) \le  P(\mbox{\boldmath $y$}| \mbox{\boldmath $x$}_{(\tilde{\mbox{\scriptsize \boldmath $w$}}, \tilde{\mbox{\scriptsize \boldmath $r$}})})\right\}, \nonumber \\
&& \qquad \qquad (\tilde{\mbox{\boldmath $w$}}, \tilde{\mbox{\boldmath $r$}}), \tilde{\mbox{\boldmath $r$}} \in \mathcal{R}, (\tilde{\mbox{\boldmath $w$}}_{\mathcal{S}}, \tilde{\mbox{\boldmath $r$}}_{\mathcal{S}})=(\mbox{\boldmath $w$}_{\mathcal{S}}, \mbox{\boldmath $r$}_{\mathcal{S}}), (\tilde{w}_k, \tilde{r}_k)\neq (w_k, r_k), \forall k\not\in \mathcal{S}.
\end{eqnarray}

Second, assume $(\mbox{\boldmath $w$}, \mbox{\boldmath $r$})$ is the transmitted message and rate pair with $\mbox{\boldmath $r$}\in \mathcal{R}$. We define $P_{t[\mbox{\scriptsize \boldmath $r$}, \mathcal{S}]}$ as the probability that the likelihood of the transmitted codeword is no larger than the predetermined threshold $\tau_{(\mbox{\scriptsize \boldmath $r$}, \mathcal{S})}(\mbox{\boldmath $y$})$.
\begin{equation}
P_{t[\mbox{\scriptsize \boldmath $r$}, \mathcal{S}]}= Pr \left \{ P(\mbox{\boldmath $y$}| \mbox{\boldmath $x$}_{(\mbox{\scriptsize \boldmath $w$}, \mbox{\scriptsize \boldmath $r$})}) \le  e^{-N\tau_{(\mbox{\tiny \boldmath $r$}, \mathcal{S})}(\mbox{\scriptsize \boldmath $y$})}\right\},
\label{PtMulti}
\end{equation}
where the threshold $\tau_{(\mbox{\scriptsize \boldmath $r$}, \mathcal{S})}(\mbox{\boldmath $y$})$ will be optimized later\footnote{As in the single-user case, the subscript $\mbox{\boldmath $r$}$ of $\tau_{(\mbox{\scriptsize \boldmath $r$}, \mathcal{S})}(\mbox{\boldmath $y$})$ represents the corresponding estimated rate of the receiver output. Note that we do not assume the receiver should know the transmitted rate.}.

Third, assume $(\tilde{\mbox{\boldmath $w$}}, \tilde{\mbox{\boldmath $r$}})$ is the transmitted message and rate pair with $\tilde{\mbox{\boldmath $r$}} \not\in \mathcal{R}$. We define $P_{i[\tilde{\mbox{\scriptsize \boldmath $r$}}, \mbox{\scriptsize \boldmath $r$}, \mathcal{S}]}$ as the probability that the receiver finds another message and rate pair $(\mbox{\boldmath $w$}, \mbox{\boldmath $r$})$ with $\mbox{\boldmath $r$} \in \mathcal{R}$, $(\mbox{\boldmath $w$}_{\mathcal{S}}, \mbox{\boldmath $r$}_{\mathcal{S}})=(\tilde{\mbox{\boldmath $w$}}_{\mathcal{S}}, \tilde{\mbox{\boldmath $r$}}_{\mathcal{S}})$, and  $(w_k, r_k)\neq (\tilde{w}_k, \tilde{r}_k), \forall k\not\in \mathcal{S}$, that has a likelihood value above the required threshold $\tau_{(\mbox{\scriptsize \boldmath $r$}, \mathcal{S})}(\mbox{\boldmath $y$})$.
\begin{eqnarray}
&& P_{i[\tilde{\mbox{\scriptsize \boldmath $r$}}, \mbox{\scriptsize \boldmath $r$}, \mathcal{S}]} = Pr\left\{ P(\mbox{\boldmath $y$}| \mbox{\boldmath $x$}_{(\mbox{\scriptsize \boldmath $w$}, \mbox{\scriptsize \boldmath $r$})})> e^{-N\tau_{(\mbox{\tiny \boldmath $r$}, \mathcal{S})}(\mbox{\scriptsize \boldmath $y$})}\right\}, \nonumber \\
&& \qquad \qquad (\mbox{\boldmath $w$}, \mbox{\boldmath $r$}), \mbox{\boldmath $r$} \in \mathcal{R}, (\mbox{\boldmath $w$}_{\mathcal{S}}, \mbox{\boldmath $r$}_{\mathcal{S}})=(\tilde{\mbox{\boldmath $w$}}_{\mathcal{S}}, \tilde{\mbox{\boldmath $r$}}_{\mathcal{S}}), (w_k, r_k)\neq (\tilde{w}_k, \tilde{r}_k), \forall k\not\in \mathcal{S}.
\end{eqnarray}

With these probability definitions, we can upper bound the system error probability $P_{es}$ by
\begin{equation}
P_{es}\le \max\left\{ \begin{array}{l} \max_{\mbox{\scriptsize \boldmath $r$}\in \mathcal{R} }\sum_{\mathcal{S}\subset\{1, \cdots, K\}} \left[ \sum_{\tilde{\mbox{\scriptsize \boldmath $r$}}\in \mathcal{R}, \tilde{\mbox{\scriptsize \boldmath $r$}}_{\mathcal{S}}= \mbox{\scriptsize \boldmath $r$}_{\mathcal{S}} } P_{m[\mbox{\scriptsize \boldmath $r$}, \tilde{\mbox{\scriptsize \boldmath $r$}}, \mathcal{S}]} +  P_{t[\mbox{\scriptsize \boldmath $r$}, \mathcal{S}]} \right],  \\  \max_{\tilde{\mbox{\scriptsize \boldmath $r$}}\not\in \mathcal{R}} \sum_{\mathcal{S}\subset\{1, \cdots, K\}} \sum_{\mbox{\scriptsize \boldmath $r$}\in \mathcal{R}, \mbox{\scriptsize \boldmath $r$}_{\mathcal{S}}=\tilde{\mbox{\scriptsize \boldmath $r$}}_{\mathcal{S}} } P_{i[\tilde{\mbox{\scriptsize \boldmath $r$}}, \mbox{\scriptsize \boldmath $r$}, \mathcal{S}]}  \end{array}     \right\}.
\label{MPesBound1}
\end{equation}
Next, we will upper bound each of the probability terms on the right hand side of (\ref{MPesBound1}).

{\bf Step 1: } Upper-bounding $P_{m[\mbox{\scriptsize \boldmath $r$}, \tilde{\mbox{\scriptsize \boldmath $r$}}, \mathcal{S}]}$.

Assume $(\mbox{\boldmath $w$}, \mbox{\boldmath $r$})$ is the transmitted message and rate pair with $\mbox{\boldmath $r$}\in \mathcal{R}$. Given $\mbox{\boldmath $r$}, \tilde{\mbox{\boldmath $r$}}\in \mathcal{R}$, $P_{m[\mbox{\scriptsize \boldmath $r$}, \tilde{\mbox{\scriptsize \boldmath $r$}}, \mathcal{S}]}$ can be written as
\begin{equation}
P_{m[\mbox{\scriptsize \boldmath $r$}, \tilde{\mbox{\scriptsize \boldmath $r$}}, \mathcal{S}]}=E_{\mbox{\scriptsize \boldmath $\theta$}}\left[\sum_{\mbox{\scriptsize \boldmath $y$} }P(\mbox{\boldmath $y$}| \mbox{\boldmath $x$}_{(\mbox{\scriptsize \boldmath $w$}, \mbox{\scriptsize \boldmath $r$})}) \phi_{m[\mbox{\scriptsize \boldmath $r$}, \tilde{\mbox{\scriptsize \boldmath $r$}}, \mathcal{S}]}( \mbox{\boldmath $y$} ) \right],
\end{equation}
where $\phi_{m[\mbox{\scriptsize \boldmath $r$}, \tilde{\mbox{\scriptsize \boldmath $r$}}, \mathcal{S}]}( \mbox{\boldmath $y$} ) =1$ if $P(\mbox{\boldmath $y$}| \mbox{\boldmath $x$}_{(\mbox{\scriptsize \boldmath $w$}, \mbox{\scriptsize \boldmath $r$})}) \le P(\mbox{\boldmath $y$}| \mbox{\boldmath $x$}_{(\tilde{\mbox{\scriptsize \boldmath $w$}}, \tilde{\mbox{\scriptsize \boldmath $r$}})})$ for some $(\tilde{\mbox{\boldmath $w$}}, \tilde{\mbox{\boldmath $r$}})$, with $(\tilde{\mbox{\boldmath $w$}}_{\mathcal{S}}, \tilde{\mbox{\boldmath $r$}}_{\mathcal{S}})=(\mbox{\boldmath $w$}_{\mathcal{S}}, \mbox{\boldmath $r$}_{\mathcal{S}})$, and $(\tilde{w}_k, \tilde{r}_k)\neq (w_k, r_k), \forall k\not\in \mathcal{S}$. $\phi_{m[\mbox{\scriptsize \boldmath $r$}, \tilde{\mbox{\scriptsize \boldmath $r$}}, \mathcal{S}]}( \mbox{\boldmath $y$} ) =0$ otherwise.

For any $\rho>0$ and $s>0$, we can bound $\phi_{m[\mbox{\scriptsize \boldmath $r$}, \tilde{\mbox{\scriptsize \boldmath $r$}}, \mathcal{S}]}( \mbox{\boldmath $y$} ) $ by
\begin{equation}
\phi_{m[\mbox{\scriptsize \boldmath $r$}, \tilde{\mbox{\scriptsize \boldmath $r$}}, \mathcal{S}]}( \mbox{\boldmath $y$} ) \le \left[ \frac{\sum_{\tilde{\mbox{\scriptsize \boldmath $w$}}, (\tilde{\mbox{\scriptsize \boldmath $w$}}_{\mathcal{S}}, \tilde{\mbox{\scriptsize \boldmath $r$}}_{\mathcal{S}})=(\mbox{\scriptsize \boldmath $w$}_{\mathcal{S}}, \mbox{\scriptsize \boldmath $r$}_{\mathcal{S}}), (\tilde{w}_k, \tilde{r}_k)\neq (w_k, r_k), \forall k\not\in \mathcal{S}}P(\mbox{\boldmath $y$}| \mbox{\boldmath $x$}_{(\tilde{\mbox{\scriptsize \boldmath $w$}}, \tilde{\mbox{\scriptsize \boldmath $r$}})})^{\frac{s}{\rho}} }{P(\mbox{\boldmath $y$}| \mbox{\boldmath $x$}_{(\mbox{\scriptsize \boldmath $w$}, \mbox{\scriptsize \boldmath $r$})})^{\frac{s}{\rho}}} \right]^{\rho}, \quad \rho>0, s>0.
\end{equation}
Consequently, $P_{m[\mbox{\scriptsize \boldmath $r$}, \tilde{\mbox{\scriptsize \boldmath $r$}}, \mathcal{S}]}$ is upper bounded by
\begin{eqnarray}
&& P_{m[\mbox{\scriptsize \boldmath $r$}, \tilde{\mbox{\scriptsize \boldmath $r$}}, \mathcal{S}]} \le E_{\mbox{\scriptsize \boldmath $\theta$}}\left[\sum_{\mbox{\scriptsize \boldmath $y$} } P(\mbox{\boldmath $y$}| \mbox{\boldmath $x$}_{(\mbox{\scriptsize \boldmath $w$}, \mbox{\scriptsize \boldmath $r$})}) \left[ \frac{\sum_{\tilde{\mbox{\scriptsize \boldmath $w$}}, (\tilde{\mbox{\scriptsize \boldmath $w$}}_{\mathcal{S}}, \tilde{\mbox{\scriptsize \boldmath $r$}}_{\mathcal{S}})=(\mbox{\scriptsize \boldmath $w$}_{\mathcal{S}}, \mbox{\scriptsize \boldmath $r$}_{\mathcal{S}}), (\tilde{w}_k, \tilde{r}_k)\neq (w_k, r_k), \forall k\not\in \mathcal{S}}P(\mbox{\boldmath $y$}| \mbox{\boldmath $x$}_{(\tilde{\mbox{\scriptsize \boldmath $w$}}, \tilde{\mbox{\scriptsize \boldmath $r$}})})^{\frac{s}{\rho}} }{P(\mbox{\boldmath $y$}| \mbox{\boldmath $x$}_{(\mbox{\scriptsize \boldmath $w$}, \mbox{\scriptsize \boldmath $r$})})^{\frac{s}{\rho}}} \right]^{\rho} \right] \nonumber \\
&& =\sum_{\mbox{\scriptsize \boldmath $y$} }  E_{\mbox{\scriptsize \boldmath $\theta$}}\left[P(\mbox{\boldmath $y$}| \mbox{\boldmath $x$}_{(\mbox{\scriptsize \boldmath $w$}, \mbox{\scriptsize \boldmath $r$})})^{1-s} \left[ \sum_{\tilde{\mbox{\scriptsize \boldmath $w$}}, (\tilde{\mbox{\scriptsize \boldmath $w$}}_{\mathcal{S}}, \tilde{\mbox{\scriptsize \boldmath $r$}}_{\mathcal{S}})=(\mbox{\scriptsize \boldmath $w$}_{\mathcal{S}}, \mbox{\scriptsize \boldmath $r$}_{\mathcal{S}}), (\tilde{w}_k, \tilde{r}_k)\neq (w_k, r_k), \forall k\not\in \mathcal{S}}P(\mbox{\boldmath $y$}| \mbox{\boldmath $x$}_{(\tilde{\mbox{\scriptsize \boldmath $w$}}, \tilde{\mbox{\scriptsize \boldmath $r$}})})^{\frac{s}{\rho}} \right]^{\rho}  \right] \nonumber \\
&& =\sum_{\mbox{\scriptsize \boldmath $y$} } E_{\mbox{\scriptsize \boldmath $\theta$}_{\mathcal{S}}}\left[  E_{\mbox{\scriptsize \boldmath $\theta$}_{\bar{\mathcal{S}}}}\left[P(\mbox{\boldmath $y$}| \mbox{\boldmath $x$}_{(\mbox{\scriptsize \boldmath $w$}, \mbox{\scriptsize \boldmath $r$})})^{1-s}\right]  E_{\mbox{\scriptsize \boldmath $\theta$}_{\bar{\mathcal{S}}}}\left[\left[ \sum_{\tilde{\mbox{\scriptsize \boldmath $w$}}, (\tilde{\mbox{\scriptsize \boldmath $w$}}_{\mathcal{S}}, \tilde{\mbox{\scriptsize \boldmath $r$}}_{\mathcal{S}})=(\mbox{\scriptsize \boldmath $w$}_{\mathcal{S}}, \mbox{\scriptsize \boldmath $r$}_{\mathcal{S}}), (\tilde{w}_k, \tilde{r}_k)\neq (w_k, r_k), \forall k\not\in \mathcal{S}}P(\mbox{\boldmath $y$}| \mbox{\boldmath $x$}_{(\tilde{\mbox{\scriptsize \boldmath $w$}}, \tilde{\mbox{\scriptsize \boldmath $r$}})})^{\frac{s}{\rho}} \right]^{\rho}  \right]\right], \nonumber \\
\label{MInequality1.1}
\end{eqnarray}
where in the last step, we can take the expectations operations over users not in $\mathcal{S}$ due to independence between the codewords of $(\mbox{\boldmath $w$}_{\bar{\mathcal{S}}}, \mbox{\boldmath $r$}_{\bar{\mathcal{S}}})$ and $(\tilde{\mbox{\boldmath $w$}}_{\bar{\mathcal{S}}}, \tilde{\mbox{\boldmath $r$}}_{\bar{\mathcal{S}}})$.

Now assume $0<\rho\le 1$. Inequality (\ref{MInequality1.1}) can be further bounded by
\begin{eqnarray}
P_{m[\mbox{\scriptsize \boldmath $r$}, \tilde{\mbox{\scriptsize \boldmath $r$}}, \mathcal{S}]}  & \le & \sum_{\mbox{\scriptsize \boldmath $y$} }  E_{\mbox{\scriptsize \boldmath $\theta$}_{\mathcal{S}}}\left[ E_{\mbox{\scriptsize \boldmath $\theta$}_{\bar{\mathcal{S}}}}\left[P(\mbox{\boldmath $y$}| \mbox{\boldmath $x$}_{(\mbox{\scriptsize \boldmath $w$}, \mbox{\scriptsize \boldmath $r$})})^{1-s}\right]  E_{\mbox{\scriptsize \boldmath $\theta$}_{\bar{\mathcal{S}}}}\left[\left[ \sum_{\tilde{\mbox{\scriptsize \boldmath $w$}}, (\tilde{\mbox{\scriptsize \boldmath $w$}}_{\mathcal{S}}, \tilde{\mbox{\scriptsize \boldmath $r$}}_{\mathcal{S}})=(\mbox{\scriptsize \boldmath $w$}_{\mathcal{S}}, \mbox{\scriptsize \boldmath $r$}_{\mathcal{S}})}P(\mbox{\boldmath $y$}| \mbox{\boldmath $x$}_{(\tilde{\mbox{\scriptsize \boldmath $w$}}, \tilde{\mbox{\scriptsize \boldmath $r$}})})^{\frac{s}{\rho}} \right]^{\rho}  \right] \right] \nonumber \\
& \le & e^{N\rho\sum_{k\not\in \mathcal{S}}\tilde{r}_k} \sum_{\mbox{\scriptsize \boldmath $y$} }   E_{\mbox{\scriptsize \boldmath $\theta$}_{\mathcal{S}}}\left[ E_{\mbox{\scriptsize \boldmath $\theta$}_{\bar{\mathcal{S}}}}\left[P(\mbox{\boldmath $y$}| \mbox{\boldmath $x$}_{(\mbox{\scriptsize \boldmath $w$}, \mbox{\scriptsize \boldmath $r$})})^{1-s}\right]  \left[  E_{\mbox{\scriptsize \boldmath $\theta$}_{\bar{\mathcal{S}}}}\left[P(\mbox{\boldmath $y$}| \mbox{\boldmath $x$}_{(\tilde{\mbox{\scriptsize \boldmath $w$}}, \tilde{\mbox{\scriptsize \boldmath $r$}})})^{\frac{s}{\rho}} \right] \right]^{\rho} \right].
\label{MInequality1.2}
\end{eqnarray}
Since (\ref{MInequality1.2}) holds for all $0< \rho\le 1$, $s>0$, and it is easy to verify that the bound becomes trivial for $s>1$, we have
\begin{equation}
P_{m[\mbox{\scriptsize \boldmath $r$}, \tilde{\mbox{\scriptsize \boldmath $r$}}, \mathcal{S}]} \le \exp \left\{-NE_m(\mathcal{S}, \tilde{\mbox{\boldmath $r$}}, \mbox{\boldmath $P$}_{\mbox{\scriptsize \boldmath $X$}|\mbox{\scriptsize \boldmath $r$}}, \mbox{\boldmath $P$}_{\mbox{\scriptsize \boldmath $X$}|\tilde{\mbox{\scriptsize \boldmath $r$}}}) \right\},
\label{MProofPmBound}
\end{equation}
where $E_m(\mathcal{S}, \tilde{\mbox{\boldmath $r$}}, \mbox{\boldmath $P$}_{\mbox{\scriptsize \boldmath $X$}|\mbox{\scriptsize \boldmath $r$}}, \mbox{\boldmath $P$}_{\mbox{\scriptsize \boldmath $X$}|\tilde{\mbox{\scriptsize \boldmath $r$}}})$ is given by
\begin{eqnarray}
&& E_m(\mathcal{S}, \tilde{\mbox{\boldmath $r$}}, \mbox{\boldmath $P$}_{\mbox{\scriptsize \boldmath $X$}|\mbox{\scriptsize \boldmath $r$}}, \mbox{\boldmath $P$}_{\mbox{\scriptsize \boldmath $X$}|\tilde{\mbox{\scriptsize \boldmath $r$}}}) = \max_{0<\rho \le 1} -\rho \sum_{k\not\in \mathcal{S}}\tilde{r}_k + \max_{0<s\le 1} -\log \sum_Y \sum_{\mbox{\scriptsize \boldmath $X$}_{\mathcal{S}}} \prod_{k\in \mathcal{S}} P_{X|r_k}(X_k)                    \nonumber \\
&& \quad \times \left(\sum_{\mbox{\scriptsize \boldmath $X$}_{\bar{\mathcal{S}}}}\prod_{k \not\in \mathcal{S}}P_{X|r_k}(X_k)P(Y|\mbox{\boldmath $X$})^{1-s}\right)  \left(\sum_{\mbox{\scriptsize \boldmath $X$}_{\bar{\mathcal{S}}}}\prod_{k \not\in \mathcal{S}}P_{X|\tilde{r}_k}(X_k)P(Y|\mbox{\boldmath $X$})^{\frac{s}{\rho}} \right)^{\rho}.
\label{MProofEmBound}
\end{eqnarray}

{\bf Step 2: } Upper-bounding $P_{t[\mbox{\scriptsize \boldmath $r$}, \mathcal{S}]}$.

Assume $(\mbox{\boldmath $w$}, \mbox{\boldmath $r$})$ is the transmitted message and rate pair with $\mbox{\boldmath $r$}\in \mathcal{R}$. Rewrite $P_{t[\mbox{\scriptsize \boldmath $r$}, \mathcal{S}]}$ as
\begin{equation}
P_{t[\mbox{\scriptsize \boldmath $r$}, \mathcal{S}]}=E_{\mbox{\scriptsize \boldmath $\theta$}}\left[\sum_{\mbox{\scriptsize \boldmath $y$} }P(\mbox{\boldmath $y$}| \mbox{\boldmath $x$}_{(\mbox{\scriptsize \boldmath $w$}, \mbox{\scriptsize \boldmath $r$})})\phi_{t[\mbox{\scriptsize \boldmath $r$}, \mathcal{S}]}(\mbox{\boldmath $y$})  \right],
\end{equation}
where $\phi_{t[\mbox{\scriptsize \boldmath $r$}, \mathcal{S}]}(\mbox{\boldmath $y$})=1$ if $P(\mbox{\boldmath $y$}| \mbox{\boldmath $x$}_{(\mbox{\scriptsize \boldmath $w$}, \mbox{\scriptsize \boldmath $r$})})\le e^{-N\tau_{(\mbox{\tiny \boldmath $r$}, \mathcal{S})}(\mbox{\scriptsize \boldmath $y$})}$, otherwise $\phi_{t[\mbox{\scriptsize \boldmath $r$}, \mathcal{S}]}(\mbox{\boldmath $y$})=0$. Note that the value of $\tau_{(\mbox{\scriptsize \boldmath $r$}, \mathcal{S})}(\mbox{\boldmath $y$})$ will be specified later.

For any $s_1>0$, we can bound $\phi_{t[\mbox{\scriptsize \boldmath $r$}, \mathcal{S}]}(\mbox{\boldmath $y$})$ as
\begin{equation}
\phi_{t[\mbox{\scriptsize \boldmath $r$}, \mathcal{S}]}(\mbox{\boldmath $y$})\le \frac{e^{-Ns_1 \tau_{(\mbox{\tiny \boldmath $r$}, \mathcal{S})}(\mbox{\scriptsize \boldmath $y$})}}{ P(\mbox{\boldmath $y$}| \mbox{\boldmath $x$}_{(\mbox{\scriptsize \boldmath $w$}, \mbox{\scriptsize \boldmath $r$})})^{s_1}} , \quad s_1>0.
\end{equation}
This yields
\begin{eqnarray}
P_{t[\mbox{\scriptsize \boldmath $r$}, \mathcal{S}]} &\le & E_{\mbox{\scriptsize \boldmath $\theta$}}\left[ \sum_{\mbox{\scriptsize \boldmath $y$} }P(\mbox{\boldmath $y$}| \mbox{\boldmath $x$}_{(\mbox{\scriptsize \boldmath $w$}, \mbox{\scriptsize \boldmath $r$})})^{1-s_1} e^{-Ns_1 \tau_{(\mbox{\tiny \boldmath $r$}, \mathcal{S})}(\mbox{\scriptsize \boldmath $y$}) } \right] \nonumber \\
&=& \sum_{\mbox{\scriptsize \boldmath $y$} } E_{\mbox{\scriptsize \boldmath $\theta$}_{\mathcal{S}}}\left[ E_{\mbox{\scriptsize \boldmath $\theta$}_{\bar{\mathcal{S}}}}\left[ P(\mbox{\boldmath $y$}| \mbox{\boldmath $x$}_{(\mbox{\scriptsize \boldmath $w$}, \mbox{\scriptsize \boldmath $r$})})^{1-s_1}\right] e^{-Ns_1 \tau_{(\mbox{\tiny \boldmath $r$}, \mathcal{S})}(\mbox{\scriptsize \boldmath $y$})}\right].
\label{MBound2}
\end{eqnarray}
We will come back to this inequality later when we optimize $\tau_{(\mbox{\scriptsize \boldmath $r$}, \mathcal{S})}(\mbox{\boldmath $y$})$.

{\bf Step 3: } Upper-bounding $P_{i[\tilde{\mbox{\scriptsize \boldmath $r$}}, \mbox{\scriptsize \boldmath $r$}, \mathcal{S}]}$.

Assume $(\tilde{\mbox{\boldmath $w$}}, \tilde{\mbox{\boldmath $r$}})$ is the transmitted message and rate pair with $\tilde{\mbox{\boldmath $r$}} \not\in \mathcal{R}$. Given $\mbox{\boldmath $r$} \in \mathcal{R}$, we first rewrite $P_{i[\tilde{\mbox{\scriptsize \boldmath $r$}}, \mbox{\scriptsize \boldmath $r$}, \mathcal{S}]}$ as
\begin{equation}
P_{i[\tilde{\mbox{\scriptsize \boldmath $r$}}, \mbox{\scriptsize \boldmath $r$}, \mathcal{S}]}=E_{\mbox{\scriptsize \boldmath $\theta$}}\left[ \sum_{\mbox{\scriptsize \boldmath $y$} }P(\mbox{\boldmath $y$}| \mbox{\boldmath $x$}_{(\tilde{\mbox{\scriptsize \boldmath $w$}}, \tilde{\scriptsize \mbox{\boldmath $r$}})})\phi_{i[\tilde{\mbox{\scriptsize \boldmath $r$}}, \mbox{\scriptsize \boldmath $r$}, \mathcal{S}]}(\mbox{\boldmath $y$} ) \right],
\end{equation}
where $\phi_{i[\tilde{\mbox{\scriptsize \boldmath $r$}}, \mbox{\scriptsize \boldmath $r$}, \mathcal{S}]}(\mbox{\boldmath $y$} )=1$ if there exists $(\mbox{\boldmath $w$}, \mbox{\boldmath $r$})$ with $\mbox{\boldmath $r$} \in \mathcal{R}$, $(\mbox{\boldmath $w$}_{\mathcal{S}}, \mbox{\boldmath $r$}_{\mathcal{S}})=(\tilde{\mbox{\boldmath $w$}}_{\mathcal{S}}, \tilde{\mbox{\boldmath $r$}}_{\mathcal{S}})$, and  $(w_k, r_k)\neq (\tilde{w}_k, \tilde{r}_k), \forall k\not\in \mathcal{S}$ to satisfy $P(\mbox{\boldmath $y$}| \mbox{\boldmath $x$}_{(\mbox{\scriptsize \boldmath $w$}, \mbox{\scriptsize \boldmath $r$})})> e^{-N\tau_{(\mbox{\tiny \boldmath $r$}, \mathcal{S})}(\mbox{\scriptsize \boldmath $y$})}$. Otherwise $\phi_{i[\tilde{\mbox{\scriptsize \boldmath $r$}}, \mbox{\scriptsize \boldmath $r$}, \mathcal{S}]}(\mbox{\boldmath $y$} )=0$.

For any $s_2>0$ and $\tilde{\rho}>0$, we can bound $\phi_{i[\tilde{\mbox{\scriptsize \boldmath $r$}}, \mbox{\scriptsize \boldmath $r$}, \mathcal{S}]}(\mbox{\boldmath $y$} )$ by

\begin{equation}
\phi_{i[\tilde{\mbox{\scriptsize \boldmath $r$}}, \mbox{\scriptsize \boldmath $r$}, \mathcal{S}]}(\mbox{\boldmath $y$} ) \le \left[ \frac{\sum_{\mbox{\scriptsize \boldmath $w$}, (\mbox{\scriptsize \boldmath $w$}_{\mathcal{S}}, \mbox{\scriptsize \boldmath $r$}_{\mathcal{S}})=(\tilde{\mbox{\scriptsize \boldmath $w$}}_{\mathcal{S}}, \tilde{\mbox{\scriptsize \boldmath $r$}}_{\mathcal{S}}), (w_k, r_k)\neq (\tilde{w}_k, \tilde{r}_k), \forall k\not\in \mathcal{S}}P(\mbox{\boldmath $y$}| \mbox{\boldmath $x$}_{(\mbox{\scriptsize \boldmath $w$}, \mbox{\scriptsize \boldmath $r$})})^{\frac{s_2}{\tilde{\rho}}}}{e^{-N\frac{s_2}{\tilde{\rho}}\tau_{(\mbox{\tiny \boldmath $r$}, \mathcal{S})}(\mbox{\scriptsize \boldmath $y$})}}  \right]^{\tilde{\rho}}, \quad s_2>0, \tilde{\rho} >0.
\end{equation}
This gives,
\begin{eqnarray}
P_{i[\tilde{\mbox{\scriptsize \boldmath $r$}}, \mbox{\scriptsize \boldmath $r$}, \mathcal{S}]} &\le & \sum_{\mbox{\scriptsize \boldmath $y$} } E_{\mbox{\scriptsize \boldmath $\theta$}}\left[ P(\mbox{\boldmath $y$}| \mbox{\boldmath $x$}_{(\tilde{\mbox{\scriptsize \boldmath $w$}}, \tilde{\mbox{\scriptsize \boldmath $r$}})})\left[ \sum_{\mbox{\scriptsize \boldmath $w$}, (\mbox{\scriptsize \boldmath $w$}_{\mathcal{S}}, \mbox{\scriptsize \boldmath $r$}_{\mathcal{S}})=(\tilde{\mbox{\scriptsize \boldmath $w$}}_{\mathcal{S}}, \tilde{\mbox{\scriptsize \boldmath $r$}}_{\mathcal{S}}), (w_k, r_k)\neq (\tilde{w}_k, \tilde{r}_k), \forall k\not\in \mathcal{S}}P(\mbox{\boldmath $y$}| \mbox{\boldmath $x$}_{(\mbox{\scriptsize \boldmath $w$}, \mbox{\scriptsize \boldmath $r$})})^{\frac{s_2}{\tilde{\rho}}}  \right]^{\tilde{\rho}} e^{Ns_2\tau_{(\mbox{\tiny \boldmath $r$}, \mathcal{S})}(\mbox{\scriptsize \boldmath $y$})} \right] \nonumber \\
& \le & \sum_{\mbox{\scriptsize \boldmath $y$} } E_{\mbox{\scriptsize \boldmath $\theta$}_{\mathcal{S}}}\left[  E_{\mbox{\scriptsize \boldmath $\theta$}_{\bar{\mathcal{S}}}}\left[P(\mbox{\boldmath $y$}| \mbox{\boldmath $x$}_{(\tilde{\mbox{\scriptsize \boldmath $w$}}, \tilde{\mbox{\scriptsize \boldmath $r$}})}) \right]E_{\mbox{\scriptsize \boldmath $\theta$}_{\bar{\mathcal{S}}}}\left[ \left[\sum_{\mbox{\scriptsize \boldmath $w$}, (\mbox{\scriptsize \boldmath $w$}_{\mathcal{S}}, \mbox{\scriptsize \boldmath $r$}_{\mathcal{S}})=(\tilde{\mbox{\scriptsize \boldmath $w$}}_{\mathcal{S}}, \tilde{\mbox{\scriptsize \boldmath $r$}}_{\mathcal{S}})}P(\mbox{\boldmath $y$}| \mbox{\boldmath $x$}_{(\mbox{\scriptsize \boldmath $w$}, \mbox{\scriptsize \boldmath $r$})})^{\frac{s_2}{\tilde{\rho}}} \right]^{\tilde{\rho}} \right] e^{Ns_2\tau_{(\mbox{\tiny \boldmath $r$}, \mathcal{S})}(\mbox{\scriptsize \boldmath $y$})} \right].
\label{MInequality1.3}
\end{eqnarray}
Note that we can separate the expectation operators in the last step due to independence between the codewords of $(\mbox{\boldmath $w$}_{\bar{\mathcal{S}}}, \mbox{\boldmath $r$}_{\bar{\mathcal{S}}})$ and $(\tilde{\mbox{\boldmath $w$}}_{\bar{\mathcal{S}}}, \tilde{\mbox{\boldmath $r$}}_{\bar{\mathcal{S}}})$.

Assume $0<\tilde{\rho}\le 1$. Inequality (\ref{MInequality1.3}) leads to
\begin{eqnarray}
&& P_{i[\tilde{\mbox{\scriptsize \boldmath $r$}}, \mbox{\scriptsize \boldmath $r$}, \mathcal{S}]}  \le \sum_{\mbox{\scriptsize \boldmath $y$} } E_{\mbox{\scriptsize \boldmath $\theta$}_{\mathcal{S}}}\left[  E_{\mbox{\scriptsize \boldmath $\theta$}_{\bar{\mathcal{S}}}}\left[P(\mbox{\boldmath $y$}| \mbox{\boldmath $x$}_{(\tilde{\mbox{\scriptsize \boldmath $w$}}, \tilde{\mbox{\scriptsize \boldmath $r$}})}) \right]\left\{ E_{\mbox{\scriptsize \boldmath $\theta$}_{\bar{\mathcal{S}}}}\left[ P(\mbox{\boldmath $y$}| \mbox{\boldmath $x$}_{(\mbox{\scriptsize \boldmath $w$}, \mbox{\scriptsize \boldmath $r$})})^{\frac{s_2}{\tilde{\rho}}} \right]\right\}^{\tilde{\rho}} e^{Ns_2\tau_{(\mbox{\tiny \boldmath $r$}, \mathcal{S})}(\mbox{\scriptsize \boldmath $y$})} e^{N\tilde{\rho}\sum_{k\not\in \mathcal{S}} r_k} \right] \nonumber \\
&& \le \max_{\mbox{\scriptsize \boldmath $r$}'\not\in\mathcal{R}, \mbox{\scriptsize \boldmath $r$}'_{\mathcal{S}}=\mbox{\scriptsize \boldmath $r$}_{\mathcal{S}} } \sum_{\mbox{\scriptsize \boldmath $y$} } E_{\mbox{\scriptsize \boldmath $\theta$}_{\mathcal{S}}}\left[  E_{\mbox{\scriptsize \boldmath $\theta$}_{\bar{\mathcal{S}}}}\left[P(\mbox{\boldmath $y$}| \mbox{\boldmath $x$}_{(\mbox{\scriptsize \boldmath $w$}', \mbox{\scriptsize \boldmath $r$}')}) \right] \left\{E_{\mbox{\scriptsize \boldmath $\theta$}_{\bar{\mathcal{S}}}}\left[ P(\mbox{\boldmath $y$}| \mbox{\boldmath $x$}_{(\mbox{\scriptsize \boldmath $w$}, \mbox{\scriptsize \boldmath $r$})})^{\frac{s_2}{\tilde{\rho}}} \right] \right\}^{\tilde{\rho}} e^{Ns_2\tau_{(\mbox{\tiny \boldmath $r$}, \mathcal{S})}(\mbox{\scriptsize \boldmath $y$})} e^{N\tilde{\rho}\sum_{k\not\in \mathcal{S}} r_k} \right].
\label{MBound3}
\end{eqnarray}
Note that the bound obtained in the last step is no longer a function of $\tilde{\mbox{\boldmath $r$}}_{\bar{\mathcal{S}}}$.

{\bf Step 4: } Choosing $\tau_{(\mbox{\scriptsize \boldmath $r$}, \mathcal{S})}(\mbox{\boldmath $y$})$.

In this step, we determine the typicality threshold $\tau_{(\mbox{\scriptsize \boldmath $r$}, \mathcal{S})}(\mbox{\boldmath $y$})$ that optimizes the bounds in (\ref{MBound2}) and (\ref{MBound3}).

Define $\tilde{\mbox{\boldmath $r$}}^*\not\in\mathcal{R} $ as
\begin{equation}
\tilde{\mbox{\boldmath $r$}}^*=\mathop{\mbox{argmax}}_{\mbox{\scriptsize \boldmath $r$}'\not\in\mathcal{R}, \mbox{\scriptsize \boldmath $r$}'_{\mathcal{S}}=\mbox{\scriptsize \boldmath $r$}_{\mathcal{S}} } \sum_{\mbox{\scriptsize \boldmath $y$} } E_{\mbox{\scriptsize \boldmath $\theta$}_{\mathcal{S}}}\left[  E_{\mbox{\scriptsize \boldmath $\theta$}_{\bar{\mathcal{S}}}}\left[P(\mbox{\boldmath $y$}| \mbox{\boldmath $x$}_{(\mbox{\scriptsize \boldmath $w$}', \mbox{\scriptsize \boldmath $r$}')}) \right] \left\{E_{\mbox{\scriptsize \boldmath $\theta$}_{\bar{\mathcal{S}}}}\left[ P(\mbox{\boldmath $y$}| \mbox{\boldmath $x$}_{(\mbox{\scriptsize \boldmath $w$}, \mbox{\scriptsize \boldmath $r$})})^{\frac{s_2}{\tilde{\rho}}} \right] \right\}^{\tilde{\rho}} e^{Ns_2\tau_{(\mbox{\tiny \boldmath $r$}, \mathcal{S})}(\mbox{\scriptsize \boldmath $y$})} e^{N\tilde{\rho}\sum_{k\not\in \mathcal{S}} r_k} \right].
\end{equation}

Given $\mbox{\boldmath $r$}\in \mathcal{R}$, $\mbox{\boldmath $y$}$, and the auxiliary variables $s_1>0$, $s_2>0$, $0<\tilde{\rho}\le 1$, we choose $\tau_{(\mbox{\scriptsize \boldmath $r$}, \mathcal{S})}(\mbox{\boldmath $y$})$ such that the following equality holds.
\begin{equation}
E_{\mbox{\scriptsize \boldmath $\theta$}_{\bar{\mathcal{S}}}}\left[ P(\mbox{\boldmath $y$}| \mbox{\boldmath $x$}_{(\mbox{\scriptsize \boldmath $w$}, \mbox{\scriptsize \boldmath $r$})})^{1-s_1}\right] e^{-Ns_1 \tau_{(\mbox{\tiny \boldmath $r$}, \mathcal{S})}(\mbox{\scriptsize \boldmath $y$}) } =
E_{\mbox{\scriptsize \boldmath $\theta$}_{\bar{\mathcal{S}}}}\left[P(\mbox{\boldmath $y$}| \mbox{\boldmath $x$}_{(\tilde{\mbox{\scriptsize \boldmath $w$}}^*, \tilde{\mbox{\scriptsize \boldmath $r$}}^*)}) \right]\left\{E_{\mbox{\scriptsize \boldmath $\theta$}_{\bar{\mathcal{S}}}}\left[P(\mbox{\boldmath $y$}| \mbox{\boldmath $x$}_{(\mbox{\scriptsize \boldmath $w$}, \mbox{\scriptsize \boldmath $r$})})^{\frac{s_2}{\tilde{\rho}}} \right] \right\}^{\tilde{\rho}} e^{Ns_2\tau_{(\mbox{\tiny \boldmath $r$}, \mathcal{S})}(\mbox{\scriptsize \boldmath $y$})} e^{N\tilde{\rho}\sum_{k\not\in \mathcal{S}} r_k}.
\label{MInequality1.5}
\end{equation}
This is always possible since the left hand side of (\ref{MInequality1.5}) decreases in $\tau_{(\mbox{\scriptsize \boldmath $r$}, \mathcal{S})}(\mbox{\boldmath $y$})$ while the right hand side of (\ref{MInequality1.5}) increases in $\tau_{(\mbox{\scriptsize \boldmath $r$}, \mathcal{S})}(\mbox{\boldmath $y$})$.

Equation (\ref{MInequality1.5}) implies
\begin{equation}
e^{-N \tau_{(\mbox{\scriptsize \boldmath $r$}, \mathcal{S})}(\mbox{\boldmath $y$}) } =\frac{ \left\{E_{\mbox{\scriptsize \boldmath $\theta$}_{\bar{\mathcal{S}}}}\left[P(\mbox{\boldmath $y$}| \mbox{\boldmath $x$}_{(\tilde{\mbox{\scriptsize \boldmath $w$}}^*, \tilde{\mbox{\scriptsize \boldmath $r$}}^*)}) \right]\right\}^{\frac{1}{s_1+s_2}}\left\{E_{\mbox{\scriptsize \boldmath $\theta$}_{\bar{\mathcal{S}}}}\left[P(\mbox{\boldmath $y$}| \mbox{\boldmath $x$}_{(\mbox{\scriptsize \boldmath $w$}, \mbox{\scriptsize \boldmath $r$})})^{\frac{s_2}{\tilde{\rho}}} \right] \right\}^{\frac{\tilde{\rho}}{s_1+s_2}} e^{N\frac{\tilde{\rho}}{s_1+s_2}\sum_{k\not\in \mathcal{S}} r_k} } { \left\{E_{\mbox{\scriptsize \boldmath $\theta$}_{\bar{\mathcal{S}}}}\left[ P(\mbox{\boldmath $y$}| \mbox{\boldmath $x$}_{(\mbox{\scriptsize \boldmath $w$}, \mbox{\scriptsize \boldmath $r$})})^{1-s_1}\right]\right\}^{\frac{1}{s_1+s_2}} }.
\label{MEquality1.6}
\end{equation}

Substitute (\ref{MEquality1.6}) into (\ref{MBound2}), we get
\begin{eqnarray}
&& P_{t[\mbox{\scriptsize \boldmath $r$}, \mathcal{S}]} \le  \sum_{\mbox{\scriptsize \boldmath $y$} } E_{\mbox{\scriptsize \boldmath $\theta$}_{\mathcal{S}}}\left[ \left\{E_{\mbox{\scriptsize \boldmath $\theta$}_{\bar{\mathcal{S}}}}\left[ P(\mbox{\boldmath $y$}| \mbox{\boldmath $x$}_{(\mbox{\scriptsize \boldmath $w$}, \mbox{\scriptsize \boldmath $r$})})^{1-s_1}\right]\right\}^{\frac{s_2}{s_1+s_2}} \left\{E_{\mbox{\scriptsize \boldmath $\theta$}_{\bar{\mathcal{S}}}}\left[P(\mbox{\boldmath $y$}| \mbox{\boldmath $x$}_{(\tilde{\mbox{\scriptsize \boldmath $w$}}^*, \tilde{\mbox{\scriptsize \boldmath $r$}}^*)}) \right]\right\}^{\frac{s_1}{s_1+s_2}} \right. \nonumber \\
&& \qquad \qquad \quad \left. \times\left\{E_{\mbox{\scriptsize \boldmath $\theta$}_{\bar{\mathcal{S}}}}\left[P(\mbox{\boldmath $y$}| \mbox{\boldmath $x$}_{(\mbox{\scriptsize \boldmath $w$}, \mbox{\scriptsize \boldmath $r$})})^{\frac{s_2}{\tilde{\rho}}} \right] \right\}^{\frac{s_1\tilde{\rho}}{s_1+s_2}} e^{N\frac{s_1\tilde{\rho}}{s_1+s_2}\sum_{k\not\in \mathcal{S}} r_k} \right].
\label{MEquality1.7}
\end{eqnarray}

Assume $s_2 < \tilde{\rho}$. Let $s_1=1-\frac{s_2}{\tilde{\rho}}$. Inequality (\ref{MEquality1.7}) becomes
\begin{equation}
P_{t[\mbox{\scriptsize \boldmath $r$}, \mathcal{S}]} \le \sum_{\mbox{\scriptsize \boldmath $y$} } E_{\mbox{\scriptsize \boldmath $\theta$}_{\mathcal{S}}}\left[ \left\{E_{\mbox{\scriptsize \boldmath $\theta$}_{\bar{\mathcal{S}}}}\left[P(\mbox{\boldmath $y$}| \mbox{\boldmath $x$}_{(\mbox{\scriptsize \boldmath $w$}, \mbox{\scriptsize \boldmath $r$})})^{\frac{s_2}{\tilde{\rho}}} \right] \right\}^{\frac{ \tilde{\rho}^2}{\tilde{\rho}-(1-\tilde{\rho})s_2}} \left\{E_{\mbox{\scriptsize \boldmath $\theta$}_{\bar{\mathcal{S}}}}\left[P(\mbox{\boldmath $y$}| \mbox{\boldmath $x$}_{(\tilde{\mbox{\scriptsize \boldmath $w$}}, \tilde{\mbox{\scriptsize \boldmath $r$}}^*)}) \right]\right\}^{\frac{\tilde{\rho}-s_2}{\tilde{\rho}-(1-\tilde{\rho})s_2}} e^{N\frac{\tilde{\rho}(\tilde{\rho}-s_2)}{\tilde{\rho}-(1-\tilde{\rho})s_2}\sum_{k\not\in \mathcal{S}} r_k} \right].
\label{MEquality1.8}
\end{equation}

Now do a variable change with $\rho=\frac{\tilde{\rho}(\tilde{\rho}-s_2)}{\tilde{\rho}-(1-\tilde{\rho})s_2}$ and $s=1-\frac{\tilde{\rho}-s_2}{\tilde{\rho}-(1-\tilde{\rho})s_2}$, and note that $s+\rho\le 1$. Inequality (\ref{MEquality1.8}) becomes
\begin{eqnarray}
&& P_{t[\mbox{\scriptsize \boldmath $r$}, \mathcal{S}]}  \le \sum_{\mbox{\scriptsize \boldmath $y$} } E_{\mbox{\scriptsize \boldmath $\theta$}_{\mathcal{S}}}\left[ \left\{E_{\mbox{\scriptsize \boldmath $\theta$}_{\bar{\mathcal{S}}}}\left[P(\mbox{\boldmath $y$}| \mbox{\boldmath $x$}_{(\mbox{\scriptsize \boldmath $w$}, \mbox{\scriptsize \boldmath $r$})})^{\frac{s}{s+\rho}}\right]\right\}^{s+\rho} \left\{E_{\mbox{\scriptsize \boldmath $\theta$}_{\bar{\mathcal{S}}}}\left[P(\mbox{\boldmath $y$}| \mbox{\boldmath $x$}_{(\tilde{\mbox{\scriptsize \boldmath $w$}}^*, \tilde{\mbox{\scriptsize \boldmath $r$}}^*)}) \right]\right\}^{1-s}e^{N\rho \sum_{k\not\in \mathcal{S}} r_k} \right]\nonumber \\
&& \le \max_{\mbox{\scriptsize \boldmath $r$}'\not \in \mathcal{R}, \mbox{\scriptsize \boldmath $r$}'_{\mathcal{S}}= \mbox{\scriptsize \boldmath $r$}_{\mathcal{S}}}\left\{ \sum_Y \sum_{\mbox{\scriptsize \boldmath $X$}_{\mathcal{S}}} \prod_{k\in \mathcal{S}} P_{X|r_k}(X_k)\left(\sum_{\mbox{\scriptsize \boldmath $X$}_{\bar{\mathcal{S}}}}\prod_{k \not\in \mathcal{S}}P_{X|r_k}(X_k)P(Y|\mbox{\boldmath $X$})^{\frac{s}{s+\rho}} \right)^{s+\rho} \right. \nonumber \\
&& \qquad \left. \times \left(\sum_{\mbox{\scriptsize \boldmath $X$}_{\bar{\mathcal{S}}}}\prod_{k \not\in \mathcal{S}}P_{X|r'_k}(X_k)P(Y|\mbox{\boldmath $X$})\right)^{1-s} \right\}^Ne^{N\rho \sum_{k\not\in \mathcal{S}} r_k}.
\label{MEquality1.9}
\end{eqnarray}

Following the same derivation, we can see that $P_{i[\tilde{\mbox{\scriptsize \boldmath $r$}}, \mbox{\scriptsize \boldmath $r$}, \mathcal{S}]}$ is also upper-bounded by the right hand side of (\ref{MEquality1.9}). Because (\ref{MEquality1.9}) holds for all $0<\rho \le 1$ and $0<s \le 1-\rho$, we have
\begin{equation}
P_{t[\mbox{\scriptsize \boldmath $r$}, \mathcal{S}]}, P_{i[\tilde{\mbox{\scriptsize \boldmath $r$}}, \mbox{\scriptsize \boldmath $r$}, \mathcal{S}]} \le \max_{\mbox{\scriptsize \boldmath $r$}'\not \in \mathcal{R}, \mbox{\scriptsize \boldmath $r$}'_{\mathcal{S}}= \mbox{\scriptsize \boldmath $r$}_{\mathcal{S}}} \exp\{ -N E_i(\mathcal{S}, \mbox{\boldmath $r$}, \mbox{\boldmath $P$}_{\mbox{\scriptsize \boldmath $X$}|\mbox{\scriptsize \boldmath $r$}}, \mbox{\boldmath $P$}_{\mbox{\scriptsize \boldmath $X$}|\mbox{\scriptsize \boldmath $r$}'})\},
\label{MProofPiBound}
\end{equation}
where
\begin{eqnarray}
&& E_i(\mathcal{S}, \mbox{\boldmath $r$}, \mbox{\boldmath $P$}_{\mbox{\scriptsize \boldmath $X$}|\mbox{\scriptsize \boldmath $r$}}, \mbox{\boldmath $P$}_{\mbox{\scriptsize \boldmath $X$}|\mbox{\scriptsize \boldmath $r$}'}) = \max_{0<\rho \le 1} -\rho \sum_{k\not\in \mathcal{S}}r_k + \max_{0<s\le 1-\rho} - \log \sum_Y \sum_{\mbox{\scriptsize \boldmath $X$}_{\mathcal{S}}} \prod_{k\in \mathcal{S}} P_{X|r_k}(X_k)               \nonumber \\
&& \quad \times \left(\sum_{\mbox{\scriptsize \boldmath $X$}_{\bar{\mathcal{S}}}}\prod_{k \not\in \mathcal{S}}P_{X|r_k}(X_k)P(Y|\mbox{\boldmath $X$})^{\frac{s}{s+\rho}} \right)^{s+\rho}\left(\sum_{\mbox{\scriptsize \boldmath $X$}_{\bar{\mathcal{S}}}}\prod_{k \not\in \mathcal{S}}P_{X|r'_k}(X_k)P(Y|\mbox{\boldmath $X$})\right)^{1-s}.
\label{MProofEiBound}
\end{eqnarray}

Finally, substituting (\ref{MProofPmBound}) and (\ref{MProofPiBound}) into (\ref{MPesBound1}) gives the desired result.

\end{proof}

\subsection{Proof of Theorem \ref{Theorem3}}
\label{ProofTheorem3}
\begin{proof}
We first present in the following lemma an achievable error probability bound for a given codeword length $N$.

\begin{lemma}{\label{Lemma1}}
Consider $K$-user random multiple access communication over a discrete-time memoryless channel $P_{Y|\mbox{\scriptsize \boldmath  $X$}}$. Assume generalized random coding $(\mbox{\boldmath  ${\cal L}$}^{(N)}, \mbox{\boldmath  $\gamma$}^{(N)})$ with a finite codeword length $N$ and $e^{NR_{\max}}$ codewords in each codebook. Let the codewords of user $k$ be partitioned into $M_k$ classes, with the $i^{th}$ codeword class corresponding to the standard rate interval $(r_{k,i-1}^U, r_{k, i}^U]$. Assume $ r_{k, 0}^U<0\le r_{k,1}^U\le r_{k,2}^U \cdots \le r_{k, M_k}^U=R_{\max}$. We term  $\{r_{k,1}^U, r_{k,2}^U, \cdots, r_{k, M_k}^U\}$ the grid rates of user $k$. For any rate $r_k \in (r_{k, i-1}^U, r_{k, i}^U]$, we define function $U(r_k)=r_{k, i}^U$, which rounds $r_k$ to its grid rate value. Let $\mbox{\boldmath  $U$}(\mbox{\boldmath  $r$})$ be the vector version of the $U(r)$ function. Denote $\mbox{\boldmath $r$}^U$ as a rate vector whose entries only take grid rate values of the corresponding users. Given an operation region $\mathcal{R}$ strictly contained in an achievable rate region, system error probability is upper-bounded by
\begin{equation}
P_{es}\le \max\left\{ \begin{array}{l} \max_{\mbox{\scriptsize \boldmath $r$}\in \mathcal{R} }\sum_{\mathcal{S}\subset\{1, \cdots, K\}} \left[ \sum_{\tilde{\mbox{\scriptsize \boldmath $r$}}^U, \tilde{\mbox{\scriptsize \boldmath $r$}}^U_{\mathcal{S}}=\mbox{\scriptsize \boldmath $U$}(\mbox{\scriptsize \boldmath $r$}_{\mathcal{S}}) } \exp\{-N \tilde{E}_m(\mathcal{S}, \tilde{\mbox{\boldmath $r$}}^U, \mbox{\boldmath $P$}_{\mbox{\scriptsize \boldmath $X$}|\mbox{\scriptsize \boldmath $r$}}, \mbox{\boldmath $P$}_{\mbox{\scriptsize \boldmath $X$}|\tilde{\mbox{\scriptsize \boldmath $r$}}, \forall \tilde{\mbox{\scriptsize \boldmath  $r$}}\in \mathcal{R}, \mbox{\scriptsize \boldmath  $U$}(\tilde{\mbox{\scriptsize \boldmath  $r$}})=\tilde{\mbox{\scriptsize \boldmath $r$}}^U, \tilde{\mbox{\scriptsize \boldmath  $r$}}_{\mathcal{S}}=\mbox{\scriptsize \boldmath  $r$}_{\mathcal{S}}} )\} \right. \\
 \qquad \left. +\max_{\mbox{\scriptsize \boldmath $r$}' \not\in \mathcal{R}, \mbox{\scriptsize \boldmath $r$}'_{\mathcal{S}}=\mbox{\scriptsize \boldmath $r$}_{\mathcal{S}}} \exp\{-N\tilde{E}_i(\mathcal{S}, \mbox{\boldmath $U$}(\mbox{\boldmath $r$}), \mbox{\boldmath $P$}_{\mbox{\scriptsize \boldmath $X$}|\hat{\mbox{\scriptsize \boldmath $r$}}, \forall \hat{\mbox{\scriptsize \boldmath $r$}}\in \mathcal{R}, \mbox{\scriptsize \boldmath $U$} (\hat{\mbox{\scriptsize \boldmath $r$}})=\mbox{\scriptsize \boldmath $U$} (\mbox{\scriptsize \boldmath $r$}), \hat{\mbox{\scriptsize \boldmath $r$}}_{\mathcal{S}}=\mbox{\scriptsize \boldmath $r$}'_{\mathcal{S}} }, \mbox{\boldmath $P$}_{\mbox{\scriptsize \boldmath $X$}|\mbox{\scriptsize \boldmath $r$}'}) \} \right],
\\  \max_{\tilde{\mbox{\scriptsize \boldmath $r$}} \not\in \mathcal{R}}\sum_{\mathcal{S}\subset\{1, \cdots, K\}} \sum_{\mbox{\scriptsize \boldmath $r$}^U, \mbox{\scriptsize \boldmath $r$}_{\mathcal{S}}^U=\mbox{\scriptsize \boldmath $U$}(\tilde{\mbox{\scriptsize \boldmath $r$}}_{\mathcal{S}})} \max_{\mbox{\scriptsize \boldmath $r$}' \not\in \mathcal{R}, \mbox{\scriptsize \boldmath $r$}'_{\mathcal{S}}=\tilde{\mbox{\scriptsize \boldmath $r$}}_{\mathcal{S}}} \\
\qquad \exp\{-N\tilde{E}_i(\mathcal{S}, \mbox{\boldmath $r$}^U, \mbox{\boldmath $P$}_{\mbox{\scriptsize \boldmath $X$}|\mbox{\scriptsize \boldmath $r$}, \forall \mbox{\scriptsize \boldmath $r$}\in \mathcal{R}, \mbox{\scriptsize \boldmath $U$} (\mbox{\scriptsize \boldmath $r$})=\mbox{\scriptsize \boldmath $r$}^U, \mbox{\scriptsize \boldmath $r$}_{\mathcal{S}}=\mbox{\scriptsize \boldmath $r$}'_{\mathcal{S}} }, \mbox{\boldmath $P$}_{\mbox{\scriptsize \boldmath $X$}|\mbox{\scriptsize \boldmath $r$}'}) \}  \end{array}     \right\},
\label{StandardRateSystemErrorBound}
\end{equation}
where exponents $\tilde{E}_m(\mathcal{S}, \tilde{\mbox{\boldmath $r$}}^U, \mbox{\boldmath $P$}_{\mbox{\scriptsize \boldmath $X$}|\mbox{\scriptsize \boldmath $r$}}, \mbox{\boldmath $P$}_{\mbox{\scriptsize \boldmath $X$}|\tilde{\mbox{\scriptsize \boldmath $r$}}, \forall \tilde{\mbox{\scriptsize \boldmath  $r$}}\in \mathcal{R}, \mbox{\scriptsize \boldmath  $U$}(\tilde{\mbox{\scriptsize \boldmath  $r$}})=\tilde{\mbox{\scriptsize \boldmath $r$}}^U, \tilde{\mbox{\scriptsize \boldmath  $r$}}_{\mathcal{S}}=\mbox{\scriptsize \boldmath  $r$}_{\mathcal{S}}} )$ and $\tilde{E}_i(\mathcal{S}, \mbox{\boldmath $r$}^U, \mbox{\boldmath $P$}_{\mbox{\scriptsize \boldmath $X$}|\mbox{\scriptsize \boldmath $r$}, \forall \mbox{\scriptsize \boldmath $r$}\in \mathcal{R}, \mbox{\scriptsize \boldmath $U$} (\mbox{\scriptsize \boldmath $r$})=\mbox{\scriptsize \boldmath $r$}^U, \mbox{\scriptsize \boldmath $r$}_{\mathcal{S}}=\mbox{\scriptsize \boldmath $r$}'_{\mathcal{S}} }, \mbox{\boldmath $P$}_{\mbox{\scriptsize \boldmath $X$}|\mbox{\scriptsize \boldmath $r$}'})$ are defined by
\begin{eqnarray}
&&\tilde{E}_m(\mathcal{S}, \tilde{\mbox{\boldmath $r$}}^U, \mbox{\boldmath $P$}_{\mbox{\scriptsize \boldmath $X$}|\mbox{\scriptsize \boldmath $r$}}, \mbox{\boldmath $P$}_{\mbox{\scriptsize \boldmath $X$}|\tilde{\mbox{\scriptsize \boldmath $r$}}, \forall \tilde{\mbox{\scriptsize \boldmath  $r$}}\in \mathcal{R}, \mbox{\scriptsize \boldmath  $U$}(\tilde{\mbox{\scriptsize \boldmath  $r$}})=\tilde{\mbox{\scriptsize \boldmath $r$}}^U, \tilde{\mbox{\scriptsize \boldmath  $r$}}_{\mathcal{S}}=\mbox{\scriptsize \boldmath  $r$}_{\mathcal{S}}} )= \max_{0<\rho \le 1} -\rho \sum_{k\not\in \mathcal{S}}\tilde{r}_k^U + \max_{0<s\le 1} -\log \sum_Y \sum_{\mbox{\scriptsize \boldmath $X$}_{\mathcal{S}}} \prod_{k\in \mathcal{S}} P_{X|r_k}(X_k)                    \nonumber \\
&& \quad \times \left(\sum_{\mbox{\scriptsize \boldmath $X$}_{\bar{\mathcal{S}}}}\prod_{k \not\in \mathcal{S}}P_{X|r_k}(X_k)P(Y|\mbox{\boldmath $X$})^{1-s}\right) \min_{\tilde{\mbox{\scriptsize \boldmath $r$}}\in \mathcal{R}, \mbox{\scriptsize \boldmath  $U$}(\tilde{\mbox{\scriptsize \boldmath  $r$}})=\tilde{\mbox{\scriptsize \boldmath $r$}}^U, \tilde{\mbox{\scriptsize \boldmath $r$}}_{\mathcal{S}}= \mbox{\scriptsize \boldmath $r$}_{\mathcal{S}}, } \left(\sum_{\mbox{\scriptsize \boldmath $X$}_{\bar{\mathcal{S}}}}\prod_{k \not\in \mathcal{S}}P_{X|\tilde{r}_k}(X_k)P(Y|\mbox{\boldmath $X$})^{\frac{s}{\rho}} \right)^{\rho},                       \nonumber \\
&& \tilde{E}_i(\mathcal{S}, \mbox{\boldmath $r$}^U, \mbox{\boldmath $P$}_{\mbox{\scriptsize \boldmath $X$}|\mbox{\scriptsize \boldmath $r$}, \forall \mbox{\scriptsize \boldmath $r$}\in \mathcal{R}, \mbox{\scriptsize \boldmath $U$} (\mbox{\scriptsize \boldmath $r$})=\mbox{\scriptsize \boldmath $r$}^U, \mbox{\scriptsize \boldmath $r$}_{\mathcal{S}}=\mbox{\scriptsize \boldmath $r$}'_{\mathcal{S}} }, \mbox{\boldmath $P$}_{\mbox{\scriptsize \boldmath $X$}|\mbox{\scriptsize \boldmath $r$}'}) = \max_{0<\rho \le 1} -\rho \sum_{k\not\in \mathcal{S}}r_k^U + \max_{0<s \le 1-\rho} - \log \sum_Y \sum_{\mbox{\scriptsize \boldmath $X$}_{\mathcal{S}}} \prod_{k\in \mathcal{S}} P_{X|r_k}(X_k)               \nonumber \\
&& \quad \times \left(\sum_{\mbox{\scriptsize \boldmath $X$}_{\bar{\mathcal{S}}}}\prod_{k \not\in \mathcal{S}}P_{X|r'_k}(X_k)P(Y|\mbox{\boldmath $X$})\right)^{1-s} \min_{ \mbox{\scriptsize \boldmath $r$} \in \mathcal{R}, \mbox{\scriptsize \boldmath $U$} (\mbox{\scriptsize \boldmath $r$})=\mbox{\scriptsize \boldmath $r$}^U, \mbox{\scriptsize \boldmath $r$}_{\mathcal{S}}=\mbox{\scriptsize \boldmath $r$}'_{\mathcal{S}} } \left(\sum_{\mbox{\scriptsize \boldmath $X$}_{\bar{\mathcal{S}}}}\prod_{k \not\in \mathcal{S}}P_{X|r_k}(X_k)P(Y|\mbox{\boldmath $X$})^{\frac{s}{s+\rho}} \right)^{s+\rho}.
\label{EmEiMultiStandard}
\end{eqnarray}
$\QED$
\end{lemma}

The proof of Lemma \ref{Lemma1} is given in Appendix \ref{ProofLemma1}.

We will now prove Theorem \ref{Theorem3} based on Lemma \ref{Lemma1}. Let the sequence of generalized random coding schemes $\{(\mbox{\boldmath  ${\cal L}$}^{(N)}, \mbox{\boldmath  $\gamma$}^{(N)})\}$ follow asymptotic input distribution $\mbox{\boldmath  $P$}_{\mbox{\scriptsize \boldmath  $X$}|\mbox{\scriptsize \boldmath  $r$}}$. Given a finite codeword length $N$, the input distribution of $(\mbox{\boldmath  ${\cal L}$}^{(N)}, \mbox{\boldmath  $\gamma$}^{(N)})$ is denoted by $\mbox{\boldmath  $P$}_{\mbox{\scriptsize \boldmath  $X$}|\mbox{\scriptsize \boldmath  $W$}^{(N)}}$. We assume convergence on the sequence of input distributions $\{\mbox{\boldmath  $P$}_{\mbox{\scriptsize \boldmath  $X$}|\mbox{\scriptsize \boldmath  $W$}^{(N)}}\}$ to its asymptotic limit $\mbox{\boldmath  $P$}_{\mbox{\scriptsize \boldmath  $X$}|\mbox{\scriptsize \boldmath  $r$}}$ is uniform\footnote{Note that $\{\mbox{\boldmath  $P$}_{\mbox{\scriptsize \boldmath  $X$}|\mbox{\scriptsize \boldmath  $W$}^{(N)}}\}$ is a {\it deterministic} sequence.}.

Assume for each user, say user $k$, we partition its codewords into $M_k$ classes, as described in Lemma \ref{Lemma1}. The $i^{th}$ codeword class corresponding to standard rate interval $(r_{k,i-1}^U, r_{k,i}^U]$. Assume $ r_{k, 0}^U<0\le r_{k,1}^U\le r_{k,2}^U \cdots \le r_{k, M_k}^U=R_{\max}$. For any rate $r_k \in (r_{k, i-1}^U, r_{k, i}^U]$, we define function $U(r_k)=r_{k, i}^U$, which rounds $r_k$ to its grid rate. Let $\mbox{\boldmath  $U$}(\mbox{\boldmath  $r$})$ be the vector version of the $U(r_k)$ function. Denote $\mbox{\boldmath $r$}^U$ as a rate vector whose entries only take grid rate values of the corresponding users. Given a finite codeword length $N$, and the operation region $\mathcal{R}$, system error probability is upper-bounded by (\ref{StandardRateSystemErrorBound}) given in Lemma \ref{Lemma1}. Let us regard the codebook partitioning as a rate partitioning, specified by $ r_{k, 0}^U<0\le r_{k,1}^U\le r_{k,2}^U \cdots \le r_{k, M_k}^U=R_{\max}$ for user $k$, $\forall k$. If we fix the rate partitioning and take the codeword length to infinity, we can lower-bound the system error exponent as
\begin{equation}
E_s\ge \min\left\{ \begin{array}{l} \min_{\mathcal{S}\subset\{1, \cdots, K\}}\min_{\mbox{\scriptsize \boldmath $r$} \in \mathcal{R}, \tilde{\mbox{\scriptsize \boldmath $r$}}^U }\tilde{E}_m(\mathcal{S}, \tilde{\mbox{\boldmath $r$}}^U, \mbox{\boldmath $P$}_{\mbox{\scriptsize \boldmath $X$}|\mbox{\scriptsize \boldmath $r$}}, \mbox{\boldmath $P$}_{\mbox{\scriptsize \boldmath $X$}|\tilde{\mbox{\scriptsize \boldmath $r$}}, \forall \tilde{\mbox{\scriptsize \boldmath  $r$}}\in \mathcal{R}, \mbox{\scriptsize \boldmath  $U$}(\tilde{\mbox{\scriptsize \boldmath  $r$}})=\tilde{\mbox{\scriptsize \boldmath $r$}}^U, \tilde{\mbox{\scriptsize \boldmath $r$}}_{\mathcal{S}}=\mbox{\scriptsize \boldmath $r$}_{\mathcal{S}}} ) ,  \\ \min_{\mathcal{S}\subset\{1, \cdots, K\}}\min_{\tilde{\mbox{\scriptsize \boldmath $r$}} \not\in \mathcal{R}, \mbox{\scriptsize \boldmath $r$}^U}  \tilde{E}_i(\mathcal{S}, \mbox{\boldmath $r$}^U, \mbox{\boldmath $P$}_{\mbox{\scriptsize \boldmath $X$}|\mbox{\scriptsize \boldmath $r$}, \forall \mbox{\scriptsize \boldmath $r$}\in \mathcal{R}, \mbox{\scriptsize \boldmath $U$} (\mbox{\scriptsize \boldmath $r$})=\mbox{\scriptsize \boldmath $r$}^U, \mbox{\scriptsize \boldmath $r$}_{\mathcal{S}}=\tilde{\mbox{\scriptsize \boldmath $r$}}_{\mathcal{S}} }, \mbox{\boldmath $P$}_{\mbox{\scriptsize \boldmath $X$}|\tilde{\mbox{\scriptsize \boldmath $r$}}}) \end{array} \right\},
\label{StandardRateErrorExponentBound}
\end{equation}
where $\tilde{E}_m(\mathcal{S}, \tilde{\mbox{\boldmath $r$}}^U, \mbox{\boldmath $P$}_{\mbox{\scriptsize \boldmath $X$}|\mbox{\scriptsize \boldmath $r$}}, \mbox{\boldmath $P$}_{\mbox{\scriptsize \boldmath $X$}|\tilde{\mbox{\scriptsize \boldmath $r$}}, \forall \tilde{\mbox{\scriptsize \boldmath  $r$}}\in \mathcal{R}, \mbox{\scriptsize \boldmath  $U$}(\tilde{\mbox{\scriptsize \boldmath  $r$}})=\tilde{\mbox{\scriptsize \boldmath $r$}}^U, \tilde{\mbox{\scriptsize \boldmath $r$}}_{\mathcal{S}}=\mbox{\scriptsize \boldmath $r$}_{\mathcal{S}}} )$ and $\tilde{E}_i(\mathcal{S}, \mbox{\boldmath $r$}^U, \mbox{\boldmath $P$}_{\mbox{\scriptsize \boldmath $X$}|\mbox{\scriptsize \boldmath $r$}, \forall \mbox{\scriptsize \boldmath $r$}\in \mathcal{R}, \mbox{\scriptsize \boldmath $U$} (\mbox{\scriptsize \boldmath $r$})=\mbox{\scriptsize \boldmath $r$}^U, \mbox{\scriptsize \boldmath $r$}_{\mathcal{S}}=\tilde{\mbox{\scriptsize \boldmath $r$}}_{\mathcal{S}} }, \mbox{\boldmath $P$}_{\mbox{\scriptsize \boldmath $X$}|\tilde{\mbox{\scriptsize \boldmath $r$}}})$ are defined in (\ref{EmEiMultiStandard}).

Define $\delta$ as the maximum width of the rate intervals.
\begin{equation}
\delta=\max_{k\in \{1, \cdots, K\}, i \in \{1, \cdots, M_k\}} r_{k, i}^U-r_{k, i-1}^U
\end{equation}
Because (\ref{StandardRateErrorExponentBound}) holds for any arbitrary rate partitioning, if we first take codeword length $N$ to infinity, and then slowly revise the rate partitioning by taking $\delta$ to zero (which means $M_k$ for all $k$ are taking to infinity), and make sure all input distributions within each rate class converge uniformly to a single asymptotic distribution, then (\ref{StandardRateErrorExponentBound}) implies (\ref{MSysErrorExponentBound}). Note that the action of ``slowly taking $\delta$ to zero" is valid since rate partitioning is only used as a tool for error exponent bound derivation. Revision on the rate partitioning does not require any change to the encoding and decoding schemes. The requirement that all input distributions within each rate class should converge uniformly as $\delta$ is taken to zero is also valid since the asymptotic input distribution function of each user is only discontinuous at a finite number of rate points.
\end{proof}

\subsection{Proof of Lemma \ref{Lemma1}}
\label{ProofLemma1}
\begin{proof}
Since the codewords in each codebook are partitioned into classes, we will prove Lemma \ref{Lemma1} by following steps similar to the proof of Theorem \ref{Theorem2}, with revisions on the bounding details due to the fact that input distributions corresponding to codewords within each class can be different. We will not repeat the proof of Theorem \ref{Theorem2}, but only explain the necessary revisions. Throughout the proof, whenever we talk about a message and rate pair $(\mbox{\boldmath $W$}, \mbox{\boldmath $r$})$, we assume $\mbox{\boldmath $r$}$ is the standard communication rate of $\mbox{\boldmath $W$}$.

We assume a similar decoding algorithm as given in (\ref{CriteriaMultiRevised}), with the second condition being revised to
\begin{equation}
\mbox{C2R: }-\frac{1}{N}\log Pr\{\mbox{\boldmath $y$}|\mbox{\boldmath $x$}_{(\mbox{\scriptsize \boldmath $W$}, \mbox{\scriptsize \boldmath $r$})}\} < \tau_{(\mbox{\scriptsize \boldmath $r$}_{\mathcal{S}}, \mbox{\scriptsize \boldmath $U$}(\mbox{\scriptsize \boldmath $r$}_{\bar{\mathcal{S}}}))}(\mbox{\boldmath $y$}).
\label{CriteriaMultiRevisedStandard}
\end{equation}
In other words, we assume the typicality threshold $\tau_{(\mbox{\scriptsize \boldmath $r$}_{\mathcal{S}}, \mbox{\scriptsize \boldmath $U$}(\mbox{\scriptsize \boldmath $r$}_{\bar{\mathcal{S}}}))}(\mbox{\boldmath $y$})$ is a function of the standard rates for users in $\mathcal{S}$ and a function of the grid rates for users not in $\mathcal{S}$.

Given a user subset $\mathcal{S}\subset \{1, \cdots, K\}$, we define the following three probability terms.

First, assume $(\mbox{\boldmath $W$}, \mbox{\boldmath $r$})$ is the transmitted message and rate pair with $\mbox{\boldmath $r$}\in \mathcal{R}$. We define $P_{m[\mbox{\scriptsize \boldmath $r$}, \tilde{\mbox{\scriptsize \boldmath $r$}}^U, \mathcal{S}]}$ as the probability that the receiver finds another codeword and rate pair $(\tilde{\mbox{\boldmath $W$}}, \tilde{\mbox{\boldmath $r$}})$ with $\tilde{\mbox{\boldmath $r$}} \in \mathcal{R}$, $\mbox{\boldmath $U$}(\tilde{\mbox{\boldmath $r$}})=\tilde{\mbox{\boldmath $r$}}^U$, $(\tilde{\mbox{\boldmath $W$}}_{\mathcal{S}}, \tilde{\mbox{\boldmath $r$}}_{\mathcal{S}})=(\mbox{\boldmath $W$}_{\mathcal{S}}, \mbox{\boldmath $r$}_{\mathcal{S}})$, and  $(\tilde{W}_k, \tilde{r}_k)\neq (W_k, r_k), \forall k\not\in \mathcal{S}$, that has a likelihood value no worse than the transmitted codeword. That is
\begin{eqnarray}
&& P_{m[\mbox{\scriptsize \boldmath $r$}, \tilde{\mbox{\scriptsize \boldmath $r$}}^U, \mathcal{S}]}= Pr\left\{ P(\mbox{\boldmath $y$}| \mbox{\boldmath $x$}_{(W, r)}) \le  P(\mbox{\boldmath $y$}| \mbox{\boldmath $x$}_{(\tilde{W}, \tilde{r})})\right\}, \nonumber \\
&& \qquad \qquad (\tilde{\mbox{\boldmath $W$}}, \tilde{\mbox{\boldmath $r$}}), \tilde{\mbox{\boldmath $r$}} \in \mathcal{R}, \mbox{\boldmath $U$}(\tilde{\mbox{\boldmath $r$}})=\tilde{\mbox{\boldmath $r$}}^U, (\tilde{\mbox{\boldmath $W$}}_{\mathcal{S}}, \tilde{\mbox{\boldmath $r$}}_{\mathcal{S}})=(\mbox{\boldmath $W$}_{\mathcal{S}}, \mbox{\boldmath $r$}_{\mathcal{S}}), (\tilde{W}_k, \tilde{r}_k)\neq (W_k, r_k), \forall k\not\in \mathcal{S}.
\end{eqnarray}

Second, assume $(\mbox{\boldmath $W$}, \mbox{\boldmath $r$})$ is the transmitted message and rate pair with $\mbox{\boldmath $r$}\in \mathcal{R}$. We define $P_{t[\mbox{\scriptsize \boldmath $r$}, \mathcal{S}]}$ as in (\ref{PtMulti}) except the typicality threshold is replaced by $\tau_{(\mbox{\scriptsize \boldmath $r$}_{\mathcal{S}}, \mbox{\scriptsize \boldmath $U$}(\mbox{\scriptsize \boldmath $r$}_{\bar{\mathcal{S}}}))}(\mbox{\boldmath $y$})$.

Third, assume $(\tilde{\mbox{\boldmath $W$}}, \tilde{\mbox{\boldmath $r$}})$ is the transmitted message and rate pair with $\tilde{\mbox{\boldmath $r$}} \not\in \mathcal{R}$. We define $P_{i[\tilde{\mbox{\scriptsize \boldmath $r$}}, \mbox{\scriptsize \boldmath $r$}^U, \mathcal{S}]}$ as the probability that the receiver finds another codeword and rate pair $(\mbox{\boldmath $W$}, \mbox{\boldmath $r$})$ with $\mbox{\boldmath $r$} \in \mathcal{R}$, $\mbox{\boldmath $U$}(\mbox{\boldmath $r$})=\mbox{\boldmath $r$}^U$, $(\mbox{\boldmath $W$}_{\mathcal{S}}, \mbox{\boldmath $r$}_{\mathcal{S}})=(\tilde{\mbox{\boldmath $W$}}_{\mathcal{S}}, \tilde{\mbox{\boldmath $r$}}_{\mathcal{S}})$, and  $(W_k, r_k)\neq (\tilde{W}_k, \tilde{r}_k), \forall k\not\in \mathcal{S}$, that has a likelihood value above the required threshold $\tau_{(\tilde{\mbox{\scriptsize \boldmath $r$}}_{\mathcal{S}}, \mbox{\scriptsize \boldmath $r$}^U_{\bar{\mathcal{S}}})}(\mbox{\boldmath $y$})$. That is
\begin{eqnarray}
&& P_{i[\tilde{\mbox{\scriptsize \boldmath $r$}}, \mbox{\scriptsize \boldmath $r$}^U, \mathcal{S}]} = Pr\left\{ P(\mbox{\boldmath $y$}| \mbox{\boldmath $x$}_{(\mbox{\scriptsize \boldmath $W$}, \mbox{\scriptsize \boldmath $r$})})> e^{-N\tau_{(\tilde{\mbox{\tiny \boldmath $r$}}_{\mathcal{S}}, \mbox{\tiny \boldmath $r$}^U_{\bar{\mathcal{S}}})}(\mbox{\scriptsize \boldmath $y$})}\right\}, \nonumber \\
&& \qquad \qquad (\mbox{\boldmath $W$}, \mbox{\boldmath $r$}), \mbox{\boldmath $r$} \in \mathcal{R}, \mbox{\boldmath $U$}(\mbox{\boldmath $r$})=\mbox{\boldmath $r$}^U, (\mbox{\boldmath $W$}_{\mathcal{S}}, \mbox{\boldmath $r$}_{\mathcal{S}})=(\tilde{\mbox{\boldmath $W$}}_{\mathcal{S}}, \tilde{\mbox{\boldmath $r$}}_{\mathcal{S}}), (W_k, r_k)\neq (\tilde{W}_k, \tilde{r}_k), \forall k\not\in \mathcal{S}.
\end{eqnarray}

With the probability definitions, we can upper bound the system error probability $P_{es}$ by
\begin{equation}
P_{es}\le \max\left\{ \begin{array}{l} \max_{\mbox{\scriptsize \boldmath $r$}\in \mathcal{R} }\sum_{\mathcal{S}\subset\{1, \cdots, K\}} \left[ \sum_{\tilde{\mbox{\scriptsize \boldmath $r$}}^U, \tilde{\mbox{\scriptsize \boldmath $r$}}^U_{\mathcal{S}}=\mbox{\scriptsize \boldmath $U$}(\mbox{\scriptsize \boldmath $r$}_{\mathcal{S}}) }P_{m[\mbox{\scriptsize \boldmath $r$}, \tilde{\mbox{\scriptsize \boldmath $r$}}^U, \mathcal{S}]} +  P_{t[\mbox{\scriptsize \boldmath $r$}, \mathcal{S}]} \right],  \\  \max_{\tilde{\mbox{\scriptsize \boldmath $r$}}\not\in \mathcal{R}}\sum_{\mathcal{S}\subset\{1, \cdots, K\}} \sum_{\mbox{\scriptsize \boldmath $r$}^U, \mbox{\scriptsize \boldmath $r$}^U_{\mathcal{S}}=\mbox{\scriptsize \boldmath $U$}(\tilde{\mbox{\scriptsize \boldmath $r$}}_{\mathcal{S}})} P_{i[\tilde{\mbox{\scriptsize \boldmath $r$}}, \mbox{\scriptsize \boldmath $r$}^U, \mathcal{S}]}  \end{array}     \right\}.
\label{MPesBound1Standard}
\end{equation}
We will then follow similar steps as in the proof of Theorem \ref{Theorem2} to upper bound each of the probability terms on the right hand side of (\ref{MPesBound1Standard}).

To upper bound $P_{m[\mbox{\scriptsize \boldmath $r$}, \tilde{\mbox{\scriptsize \boldmath $r$}}^U, \mathcal{S}]}$, we assume $0<\rho\le 1$, $0<s\le 1$, and get from (\ref{MInequality1.2}) that
\begin{eqnarray}
&& P_{m[\mbox{\scriptsize \boldmath $r$}, \tilde{\mbox{\scriptsize \boldmath $r$}}^U, \mathcal{S}]}   \le  \sum_{\mbox{\scriptsize \boldmath $y$} }  E_{\mbox{\scriptsize \boldmath $\theta$}_{\mathcal{S}}}\left[ E_{\mbox{\scriptsize \boldmath $\theta$}_{\bar{\mathcal{S}}}}\left[P(\mbox{\boldmath $y$}| \mbox{\boldmath $x$}_{(\mbox{\scriptsize \boldmath $W$}, \mbox{\scriptsize \boldmath $r$})})^{1-s}\right] \left[ \sum_{\tilde{\mbox{\scriptsize \boldmath $W$}}, (\tilde{\mbox{\scriptsize \boldmath $W$}}_{\mathcal{S}}, \tilde{\mbox{\scriptsize \boldmath $r$}}_{\mathcal{S}})=(\mbox{\scriptsize \boldmath $W$}_{\mathcal{S}}, \mbox{\scriptsize \boldmath $r$}_{\mathcal{S}}), \mbox{\scriptsize \boldmath $U$}(\tilde{\mbox{\scriptsize \boldmath $r$}})=\tilde{\mbox{\scriptsize \boldmath $r$}}^U}  E_{\mbox{\scriptsize \boldmath $\theta$}_{\bar{\mathcal{S}}}}\left[P(\mbox{\boldmath $y$}| \mbox{\boldmath $x$}_{(\tilde{\mbox{\scriptsize \boldmath $W$}}, \tilde{\mbox{\scriptsize \boldmath $r$}})})^{\frac{s}{\rho}} \right]^{\rho}  \right] \right] \nonumber \\
&& \le e^{N\rho\sum_{k\not\in \mathcal{S}}\tilde{r}_k^U} \sum_{\mbox{\scriptsize \boldmath $y$} }   E_{\mbox{\scriptsize \boldmath $\theta$}_{\mathcal{S}}}\left[ E_{\mbox{\scriptsize \boldmath $\theta$}_{\bar{\mathcal{S}}}}\left[P(\mbox{\boldmath $y$}| \mbox{\boldmath $x$}_{(\mbox{\scriptsize \boldmath $W$}, \mbox{\scriptsize \boldmath $r$})})^{1-s}\right]  \left[  \max_{\tilde{\mbox{\scriptsize \boldmath $W$}}, (\tilde{\mbox{\scriptsize \boldmath $W$}}_{\mathcal{S}}, \tilde{\mbox{\scriptsize \boldmath $r$}}_{\mathcal{S}})=(\mbox{\scriptsize \boldmath $W$}_{\mathcal{S}}, \mbox{\scriptsize \boldmath $r$}_{\mathcal{S}}), \mbox{\scriptsize \boldmath $U$}(\tilde{\mbox{\scriptsize \boldmath $r$}})=\tilde{\mbox{\scriptsize \boldmath $r$}}^U}E_{\mbox{\scriptsize \boldmath $\theta$}_{\bar{\mathcal{S}}}}\left[ P(\mbox{\boldmath $y$}| \mbox{\boldmath $x$}_{(\tilde{\mbox{\scriptsize \boldmath $W$}}, \tilde{\mbox{\scriptsize \boldmath $r$}})})^{\frac{s}{\rho}} \right] \right]^{\rho} \right] \nonumber \\
&& \le \exp\{-N \tilde{E}_m(\mathcal{S}, \tilde{\mbox{\boldmath $r$}}^U, \mbox{\boldmath $P$}_{\mbox{\scriptsize \boldmath $X$}|\mbox{\scriptsize \boldmath $r$}}, \mbox{\boldmath $P$}_{\mbox{\scriptsize \boldmath $X$}|\tilde{\mbox{\scriptsize \boldmath $r$}}, \forall \tilde{\mbox{\scriptsize \boldmath  $r$}}\in \mathcal{R}, \mbox{\scriptsize \boldmath  $U$}(\tilde{\mbox{\scriptsize \boldmath  $r$}})=\tilde{\mbox{\scriptsize \boldmath $r$}}^U, \tilde{\mbox{\scriptsize \boldmath $r$}}_{\mathcal{S}}=\mbox{\scriptsize \boldmath $r$}_{\mathcal{S}}} )\},
\label{MInequality1.2Standard}
\end{eqnarray}
where $\tilde{E}_m(\mathcal{S}, \tilde{\mbox{\boldmath $r$}}^U, \mbox{\boldmath $P$}_{\mbox{\scriptsize \boldmath $X$}|\mbox{\scriptsize \boldmath $r$}}, \mbox{\boldmath $P$}_{\mbox{\scriptsize \boldmath $X$}|\tilde{\mbox{\scriptsize \boldmath $r$}}, \forall \tilde{\mbox{\scriptsize \boldmath  $r$}}\in \mathcal{R}, \mbox{\scriptsize \boldmath  $U$}(\tilde{\mbox{\scriptsize \boldmath  $r$}})=\tilde{\mbox{\scriptsize \boldmath $r$}}^U, \tilde{\mbox{\scriptsize \boldmath $r$}}_{\mathcal{S}}=\mbox{\scriptsize \boldmath $r$}_{\mathcal{S}}} )$ is defined in (\ref{EmEiMultiStandard}).

To upper bound $P_{t[\mbox{\scriptsize \boldmath $r$}, \mathcal{S}]}$, we get from (\ref{MBound2}) for $s_1>0$ that
\begin{eqnarray}
&&  P_{t[\mbox{\scriptsize \boldmath $r$}, \mathcal{S}]} \le  \sum_{\mbox{\scriptsize \boldmath $y$} } E_{\mbox{\scriptsize \boldmath $\theta$}_{\mathcal{S}}}\left[ E_{\mbox{\scriptsize \boldmath $\theta$}_{\bar{\mathcal{S}}}}\left[ P(\mbox{\boldmath $y$}| \mbox{\boldmath $x$}_{(\mbox{\scriptsize \boldmath $W$}, \mbox{\scriptsize \boldmath $r$})})^{1-s_1}\right] e^{-Ns_1 \tau_{(\mbox{\tiny \boldmath $r$}_{\mathcal{S}}, \mbox{\tiny \boldmath $U$}(\mbox{\tiny \boldmath $r$}_{\bar{\mathcal{S}}}))}(\mbox{\scriptsize \boldmath $y$})}\right] .
\label{MBound2Standard}
\end{eqnarray}

To upper bound $P_{i[\tilde{\mbox{\scriptsize \boldmath $r$}}, \mbox{\scriptsize \boldmath $r$}^U, \mathcal{S}]}$, we get from (\ref{MBound3}) for $s_2>0$ and $0<\tilde{\rho}\le 1$ that
\begin{eqnarray}
&& P_{i[\tilde{\mbox{\scriptsize \boldmath $r$}}, \mbox{\scriptsize \boldmath $r$}^U, \mathcal{S}]} \le  \sum_{\mbox{\scriptsize \boldmath $y$} } E_{\mbox{\scriptsize \boldmath $\theta$}_{\mathcal{S}}}\left[  E_{\mbox{\scriptsize \boldmath $\theta$}_{\bar{\mathcal{S}}}}\left[P(\mbox{\boldmath $y$}| \mbox{\boldmath $x$}_{(\tilde{\mbox{\scriptsize \boldmath $W$}}, \tilde{\mbox{\scriptsize \boldmath $r$}})}) \right]\left\{\sum_{(\mbox{\scriptsize \boldmath $W$}, \mbox{\scriptsize \boldmath $r$}), \mbox{\scriptsize \boldmath $r$}_{\mathcal{S}}=\tilde{\mbox{\scriptsize \boldmath $r$}}_{\mathcal{S}}, \mbox{\scriptsize \boldmath $U$}(\mbox{\scriptsize \boldmath $r$}_{\bar{\mathcal{S}}})=\mbox{\scriptsize \boldmath $r$}_{\bar{\mathcal{S}}}^U } E_{\mbox{\scriptsize \boldmath $\theta$}_{\bar{\mathcal{S}}}}\left[P(\mbox{\boldmath $y$}| \mbox{\boldmath $x$}_{(\mbox{\scriptsize \boldmath $W$}, \mbox{\scriptsize \boldmath $r$})})^{\frac{s_2}{\tilde{\rho}}}\right]\right\}^{\tilde{\rho}} \right] \nonumber \\
&& \qquad \times e^{Ns_2\tau_{(\tilde{\mbox{\tiny \boldmath $r$}}_{\mathcal{S}}, \mbox{\tiny \boldmath $r$}^U_{\bar{\mathcal{S}}} )}(\mbox{\scriptsize \boldmath $y$})}\nonumber \\
&& \le \sum_{\mbox{\scriptsize \boldmath $y$} } E_{\mbox{\scriptsize \boldmath $\theta$}_{\mathcal{S}}}\left[  E_{\mbox{\scriptsize \boldmath $\theta$}_{\bar{\mathcal{S}}}}\left[P(\mbox{\boldmath $y$}| \mbox{\boldmath $x$}_{(\tilde{\mbox{\scriptsize \boldmath $W$}}, \tilde{\mbox{\scriptsize \boldmath $r$}})}) \right]\left\{\max_{(\mbox{\scriptsize \boldmath $W$}, \mbox{\scriptsize \boldmath $r$}), \mbox{\scriptsize \boldmath $r$}_{\mathcal{S}}=\tilde{\mbox{\scriptsize \boldmath $r$}}_{\mathcal{S}}, \mbox{\scriptsize \boldmath $U$}(\mbox{\scriptsize \boldmath $r$}_{\bar{\mathcal{S}}})=\mbox{\scriptsize \boldmath $r$}_{\bar{\mathcal{S}}}^U } E_{\mbox{\scriptsize \boldmath $\theta$}_{\bar{\mathcal{S}}}}\left[P(\mbox{\boldmath $y$}| \mbox{\boldmath $x$}_{(\mbox{\scriptsize \boldmath $W$}, \mbox{\scriptsize \boldmath $r$})})^{\frac{s_2}{\tilde{\rho}}}\right]\right\}^{\tilde{\rho}} \right] \nonumber \\
&& \qquad \times e^{Ns_2\tau_{(\tilde{\mbox{\tiny \boldmath $r$}}_{\mathcal{S}}, \mbox{\tiny \boldmath $r$}^U_{\bar{\mathcal{S}}}) }(\mbox{\scriptsize \boldmath $y$})}e^{N\tilde{\rho}\sum_{k\not\in \mathcal{S}}r_k^U} \nonumber \\
&& \le \max_{\mbox{\scriptsize \boldmath $r$}'\not\in \mathcal{R}, \mbox{\scriptsize \boldmath $r$}'_{\mathcal{S}}=\tilde{\mbox{\scriptsize \boldmath $r$}}_{\mathcal{S}}}
\sum_{\mbox{\scriptsize \boldmath $y$} } E_{\mbox{\scriptsize \boldmath $\theta$}_{\mathcal{S}}}\left[  E_{\mbox{\scriptsize \boldmath $\theta$}_{\bar{\mathcal{S}}}}\left[P(\mbox{\boldmath $y$}| \mbox{\boldmath $x$}_{(\mbox{\scriptsize \boldmath $W$}', \mbox{\scriptsize \boldmath $r$}')}) \right]\left\{\max_{(\mbox{\scriptsize \boldmath $W$}, \mbox{\scriptsize \boldmath $r$}), \mbox{\scriptsize \boldmath $r$}_{\mathcal{S}}=\tilde{\mbox{\scriptsize \boldmath $r$}}_{\mathcal{S}}, \mbox{\scriptsize \boldmath $U$}(\mbox{\scriptsize \boldmath $r$}_{\bar{\mathcal{S}}})=\mbox{\scriptsize \boldmath $r$}_{\bar{\mathcal{S}}}^U } E_{\mbox{\scriptsize \boldmath $\theta$}_{\bar{\mathcal{S}}}}\left[P(\mbox{\boldmath $y$}| \mbox{\boldmath $x$}_{(\mbox{\scriptsize \boldmath $W$}, \mbox{\scriptsize \boldmath $r$})})^{\frac{s_2}{\tilde{\rho}}}\right]\right\}^{\tilde{\rho}} \right] \nonumber \\
&& \qquad \times e^{Ns_2\tau_{(\tilde{\mbox{\tiny \boldmath $r$}}_{\mathcal{S}}, \mbox{\tiny \boldmath $r$}^U_{\bar{\mathcal{S}}} )}(\mbox{\scriptsize \boldmath $y$})}e^{N\tilde{\rho}\sum_{k\not\in \mathcal{S}}r_k^U}.
\label{MInequality1.3Standard}
\end{eqnarray}

Next, by following a derivation similar to Step 4 in the proof of Theorem \ref{Theorem2}, we can optimize (\ref{MBound2Standard}) and (\ref{MInequality1.3Standard}) jointly over $\tau_{(\tilde{\mbox{\scriptsize \boldmath $r$}}_{\mathcal{S}}, \mbox{\scriptsize \boldmath $r$}^U_{\bar{\mathcal{S}}} )}(\mbox{\boldmath $y$})$ to obtain the desired result.

\end{proof}



%




\end{document}